\documentclass[twocolumn,times]{aastex631} 

\usepackage{amsmath,amssymb}
\usepackage{multirow}
\usepackage{xcolor}
\usepackage{natbib}
\usepackage{longtable}
\usepackage{refcount}
\usepackage{array}
\usepackage[figuresright]{rotating}
\usepackage{afterpage}
\usepackage{fancyhdr}
\usepackage{flushend} 

\usepackage{amsmath}
\usepackage{amssymb}
\usepackage{ulem}
\usepackage{float}
\hypersetup{
     colorlinks   = true,
     citecolor    = blue,
     linkcolor=  blue,
     urlcolor=  blue
}

\makeatletter
\patchcmd{\@bibitem}{\ignorespaces}{\label{bib-#1}\ignorespaces}{}{}

\afterpage{\global\setlength\@fpsep{8\p@ \@plus 2fil}}

\patchcmd{\NAT@citex}
  {\@citea\NAT@hyper@{%
     \NAT@nmfmt{\NAT@nm}%
     \hyper@natlinkbreak{\NAT@aysep\NAT@spacechar}{\@citeb\@extra@b@citeb}%
     \NAT@date}}
  {\@citea\NAT@nmfmt{\NAT@nm}%
   \NAT@aysep\NAT@spacechar\NAT@hyper@{\NAT@date}}{}{}

\patchcmd{\NAT@citex}
  {\@citea\NAT@hyper@{%
     \NAT@nmfmt{\NAT@nm}%
     \hyper@natlinkbreak{\NAT@spacechar\NAT@@open\if*#1*\else#1\NAT@spacechar\fi}%
       {\@citeb\@extra@b@citeb}%
     \NAT@date}}
  {\@citea\NAT@nmfmt{\NAT@nm}%
   \NAT@spacechar\NAT@@open\if*#1*\else#1\NAT@spacechar\fi\NAT@hyper@{\NAT@date}}
  {}{}

\newcommand\@tmp{}
\newcommand\@refname{}
\newcommand\@refnames{}
\newcommand\@nextref{}

\newcommand\aref[1]{\@first@ref#1,@}
\def\@throw@dot#1.#2@{#1}
\def\@set@refname#1{
    \edef\@tmp{\getrefbykeydefault{#1}{anchor}{}}%
    \def\@refname{\@nameuse{\expandafter\@throw@dot\@tmp.@autorefname}}%
    \def\@refnames{\@nameuse{\expandafter\@throw@dot\@tmp.@autorefname}s}%
}
\def\@first@ref#1,#2{%
    \@set@refname{#1}
  \ifx#2@\let\@nextref\@gobble
    \@refname~\ref{#1}
  \else%
    \@refnames~\ref{#1}
    \let\@nextref\@next@ref
  \fi%
  \@nextref#2%
}
\def\@next@ref#1,#2{%
   \ifx#2@ and~\ref{#1}\let\@nextref\@gobble
   \else, \ref{#1}
   \fi%
   \@nextref#2%
}

\makeatother

\begin{document}               

\makeatother

\title{A Surprising Lack of Metallicity Evolution with Redshift in the Long Gamma-Ray Burst Host Galaxy Population}
\newcommand\pagetitle{{Lack of LGRB Host Galaxy Metallicity Evolution}} 
\newcommand\LastPage{\pageref{metal_comp_table}}

\author{J. F. Graham}\affiliation{Kavli Institute for Astronomy and Astrophysics at Peking University, No.~5 Yiheyuan Road, Haidian District, Beijing, P. R. China}
\author{P. Schady}\affiliation{Department of Physics, University of Bath, Bath BA2 7AY, United Kingdom}\affiliation{Max-Planck Institut f\"ur Extraterrestrische Physik, Giessenbachstrasse 1, 85748, Garching, Germany}
\author{A. S. Fruchter}\affiliation{Space Telescope Science Institute, 3700 San Martin Drive, Baltimore MD 21218}

\journalinfo{}

\begin{abstract}
The number of long-duration Gamma Ray Burst (LGRB) host galaxies with measured metallicities and host masses has now grown to over one hundred, allowing us to investigate how the distributions of both these properties change with redshift. Using the combined host galaxy metallicity sample from \cite{stats_paper} and \cite{xshooter_survey}, we find a surprising lack of evolution in the LGRB metallicity distribution across different redshifts. In particular, the fraction of LGRB hosts with relatively high metallicity (12+log(O/H) $\geq 8.4$) remains essentially constant out to $z=2.5$. This result is at odds with the evolution in the mass-metallicity relation of typical galaxies, which become progressively more metal poor with increasing redshift. A similar result is found when converting the LGRB host galaxy mass distribution taken from the SHOALS ({\em Swift} GRB Host Galaxy Legacy Survey) sample to a corresponding metallicity distribution by applying a redshift-dependent mass-metallicity relation. The SHOALS sample is compiled using an unbiased selection function implying that the observed lack of evolution in the host galaxy high metallicity distribution is not caused by selection effects. However, the LGRB host galaxy metallicities estimated from the stellar mass are typically a quarter dex higher at all redshifts than the metallicity we measure spectroscopically. This implies that using mass-metallicity relationships to estimate host metallicities will thus produce a substantial systematic bias.
\end{abstract}

\defcitealias{KobulnickyKewley}{KK04}
\defcitealias{Curti2017}{CCM+17}

\section{Introduction}

Upon collecting the first samples of the galaxies hosting Long-soft Gamma-Ray Bursts (LGRBs) it was apparent that LGRBs occur in blue, highly starforming, and often irregular galaxies with a preponderance that clearly separated them from the general galaxy population \citep{Fruchter1999, Fruchter, LeFlochblue, LeFlochblue2002, Christensen, LeFloch2006, Savaglio}.  This was in line with the predictions from the collapsar model, whereby LGRBs are produced by the core-collapse of rapidly rotating, sub-solar metallicity stars \citep{Woosley1993, Woosley1999, Hirschi2005,Yoon2005}.

The seminal work of \cite{Fruchter} compared the hosts of LGRBs with those of Core-Collapse Supernovae (CCSNe) found in the Great Observatories Origins Deep Survey (GOODS), and found that while half of the GOODs CCSNe occurred in grand design spirals (with the other half in irregulars), only one out of the 18 LGRB host galaxies of a comparable redshift distribution was in a grand design spiral. As the Initial Mass Function (IMF) of blue irregular and spiral galaxies are thought to be largely similar \citep{Bastian}, massive stellar progenitors should be just as available per unit star-formation in both galaxy types.  However, the much smaller size of blue irregulars would suggest, due to the galaxy mass-metallicity relation  \citep{Tremonti2004}, that blue irregular galaxies are typically metal poor in comparison with grand design spirals. In line with this, \cite{Stanek2007} showed that the very nearest LGRB hosts all have low metallicity when compared to similar magnitude galaxies in the Sloan Digital Sky Survey (SDSS), and \cite{Kewley2007} found the LGRB host sample to be comparable to extremely metal-poor galaxies in luminosity-metallicity relation, star-formation rate (SFR), and internal extinction. 
To better understand the nature of the metallicity dependence, \cite{stats_paper} compared the metallicity distribution of the hosts of relatively nearby LGRBs (z$<0.83$) with that of the hosts of several similar indicators of star-formation: LGRBs, Type Ic-bl, and Type II SNe as well as with the metallicity distribution of star-formation in the general galaxy population of the local universe.  \cite{stats_paper} found that three quarters of the LGRB host population have metallicities below 12+log(O/H)~$<$~8.6 (in the \citealt{KobulnickyKewley} KK04 metallicity scale), while less than a tenth of local star-formation is at similarly low metallicities.  However, the Type II supernovae were statistically consistent with the metallicity distribution of star-formation in the general galaxy population. The LGRB sample extends to higher redshifts than the other populations, out to cosmic epochs where galaxies was generally more metal-poor. However, the observed metallicity difference between LGRB and Type II SNe was far too great to be a product of the cosmic chemical evolution alone. \cite{stats_paper} concluded that a low metallicity environment must be a fundamental component of the evolutionary process that forms LGRBs. This analysis was further extended in \cite{diff_rate_letter}, in which they compared the metallicity distribution of LGRB host galaxies to that of star forming galaxies in the local universe, and found a dramatic cutoff in the LGRB formation rate per unit star-formation above a metallicity of log(O/H)+12 $\approx 8.3 $ (in the \citealt{KobulnickyKewley} scale), with LGRBs forming between ten and fifty times more frequently per unit star-formation below this cutoff than above.

The dramatic metallicity difference between both the star-formation weighted SDSS and non-engine driven SNe verses LGRB samples (including normal Type Ic, c.f. \citealt{Modjaz2019}) suggests a metallicity dependence in either the formation of the gamma-ray jet or in its ability to escape the progenitor, which has either burned or lost its outer hydrogen and helium layers \citep{Woosley1993, Woosley, Langer}. Nevertheless, as samples have become larger, and selection effects have been better addressed, a surprising number of LGRBs have been located in host galaxies with a high metallicity (near-solar and above) \citep[e.g.,][]{Prochaska2009,conference_proceedings,Savaglio2012,Svensson2012,Elliott2013,Perley_dusty_LGRBs,Schady2015}.  This metal-rich host galaxy population may be compatible with the collapsar model if the LGRBs occurred in low metallicity pockets of their host galaxy \citep{Metha}, or it may imply that LGRB progenitors can maintain their high angular momentum at the point of core collapse due to interactions with a binary companion, which is a less metal-sensitive channel of LGRB formation \citep{Chrimes2020}.

\cite{xshooter_survey} analyzed a large sample of 96 LGRB host galaxies ($0.1<$z$<3.6$) with ESO Very Large Telescope (VLT) X-Shooter emission-line spectroscopy, comprising the largest LGRB host galaxy spectroscopic sample to date, and although they also found that LGRBs arose preferentially in low-metallicity host galaxies, $\sim$20\% of their sample at $z<1$ had super-solar metallicities in their adopted metallicity scale. In an attempt to reduce selection biases, \cite{Palmerio2019} used the unbiased BAT6 \cite{BAT6} parent sample of LGRBs to investigate the metallicity distribution of LGRB host galaxies relative to the typical star forming galaxies from the MOSDEF \citep{Kriek2015} and COSMOS2015 \citep{Laigle2016} surveys, and similarly found a preference for LGRBs to select metal-poor galaxies, with a metallicity cut off of 12+log(O/H)=8.55. However, they also found up to 20\% and 10\% of LGRB host galaxies to have super-solar metallicities at $z<1$ and $1<z<2$, respectively. Nevertheless, the need for galaxy spectra to measure the metallicity does introduce biases to the sample, with a subsequent completeness of just $\sim$20\%.

Given the difficulties in obtaining the necessary high quality galaxy spectra to measure the metallicities of an unbiased sample of LGRB host galaxies, \cite{Vergani2015} instead  estimated the host stellar masses of the 14 $z$~$<$~1 BAT6 \citep{BAT6} host galaxies via spectral energy distribution (SED) fitting and found that those LGRBs tend to avoid massive galaxies in preference for faint low-mass star-forming galaxies typically below galaxy survey completeness limits. \cite{Vergani2015} finds that when assuming galaxies follow the fundamental metallicity relation (FMR\footnote{a redshift-invariant scaling relation between SFR, M$_\star$ and $12+\log(\rm{O/H})$) from \cite{Mannucci}, LGRB host galaxies require a metallicity upper limit of between $0.3-0.5$Z$_\odot$ to reproduce the observed LGRB host galaxy mass distribution.}

In a similar analysis, \cite{Perley_shoals_masses} used Spitzer rest-frame near-IR (NIR) luminosity observations to calculate masses for 82 LGRB host galaxies from the Swift GRB Host Galaxy Legacy Survey (SHOALS), spanning $0.03$~$<$~$z$~$<$~$5.3$, and also use the distribution of these masses to estimate a metallicity threshold.  However \cite{Perley_shoals_masses} estimates the metallicity threshold to be much higher at ``approximately the solar value."

An alternative and relatively model-independent method of probing the metallicity of GRB host galaxies is from the hydrogen and metal absorption lines imprinted on the LGRB afterglow spectrum from host galaxy intervening material \citep{Savaglio2006,Thoene2013,Wiseman,DeCia2018,Bolmer2019}. Such absorption-line metallicities can only be measured in LGRBs at $z$~$>$~1.7, at which point Ly-$\alpha$ absorption becomes accessible from the ground, but the sensitivity of absorption line spectroscopy enables metallicities to be measured out to $z\sim 6.3$ \citep{Kawai}. The distribution of absorption based metallicities are generally far lower than emission-line based metallicities, although these offsets are likely to be affected by strong selection effects, whereby sensitive, absorption lines are more easily attainable along dust-free sightlines, corresponding to more metal-poor host galaxies \citep{Schady_RSOS, DeCia2018}. Indeed, {\em dark} LGRBs \citep{Jakobssondark}, which have their optical afterglow attenuated, typically by dust, have stronger metal absorption lines in their afterglow spectra compared to optically-bright LGRBs \citep{Christensen11}.

In light of the ongoing discrepancies in the literature on the distribution of LGRB host galaxy metallicities, in this paper we exploit the large spectroscopic samples now available from \cite{xshooter_survey} and \cite{stats_paper} to examine in detail how the LGRB metallicity distribution changes with redshift. In \aref{meths}, we introduce our sample and describe our method to measure the host galaxy metallicity. Our results are presented in \aref{mzd}, where we also compare our spectroscopically-derived LGRB metallicity distributions to the expected metallicity distribution for galaxies with the mass and redshift of the LGRB population. Finally, we discuss the implications and possible interpretations of our results in \aref{Discussion}.

\section{Metallicity sample}\label{meths}
\subsection{Sample selection}\label{sample}
The vast majority of LGRB host galaxy emission-line metallicity measurements are contained in either the \cite{stats_paper} sample or the X-Shooter observations compiled in \cite{xshooter_survey}. \cite{xshooter_survey} published the X-shooter line emission data for a sample of 96 LGRB host galaxies, of which 42 have reported fluxes in the lines necessary to apply the R$_{23}$ metallicity diagnostic (i.e.\ [\ion{O}{2}]$\lambda$3727, [\ion{O}{3}]$\lambda$5007, [\ion{N}{2}]$\lambda6584$, H$\alpha$, and H$\beta$; see \aref{Host_Metallicities}). We combine this sample with the sample from \cite{stats_paper}, removing two duplications (GRBs 050824 and 051022), as well as GRB~020819B, which has a revised host galaxy association \citep{Perley020819B}, leaving us with a combined sample of 53 objects. The line fluxes in \cite{stats_paper} and \cite{xshooter_survey} for the two duplicate LGRBs (GRB~050824 and GRB~051022) are broadly consistent within $1\sigma$, although the [\ion{O}{2}]$\lambda$3727 line flux is only consistent at the $2\sigma$ and $3\sigma$ level, respectively. The effect on the subsequent measured R$_{23}$ metallicities is minimal in the case of GRB~050824 (consistent within $1\sigma$), but more notable for GRB~051022, where the agreement is at the $3\sigma$ level. Nevertheless, these differences would be smaller if uncertainties on the line measurements reported in \cite{stats_paper} were available. When removing duplications from our combined sample, we favor the flux measurements from \cite{xshooter_survey}\phantomsection\label{dupout}, because these line flux values have reported uncertainties whereas the LGRB 050824 line fluxes of \cite{Sollerman2007} (used in \citealt{stats_paper}) and the LGRB 051022 line fluxes of \cite{Levesque2} do not.  Furthermore the \cite{stats_paper} metallicity for LGRB 051022 was determined from equivalent width observations \citep{conference_proceedings}, not line flux values, which while an established technique \citep{KobulnickyPhillips}, introduces a difference in method that we can easily avoid here.

There were a further 10 LGRB host galaxies with X-shooter observations that, although analyzed, didn't make it into the final \cite{xshooter_survey} publication (private communication). Of these, two \citep[GRB~000210 and GRB~011211; see also][]{Piranomonte2015}, had the relevant emission line detections required by our selection criteria (see above), bringing our final sample to a total of 55 LGRB host galaxies. The X-shooter data for these two LGRB host galaxies were reduced and analyzed following the same method as \cite{xshooter_survey}, and we list the measured line fluxes in \aref{xsd_line_fluxes}.

\newcolumntype{L}[1]{>{\centering\arraybackslash}m{#1}}
\begin{table*}[t]
\begin{center}
\begin{minipage}[H]{1\textwidth}
\caption{\label{xsd_line_fluxes} Measured X-shooter line fluxes for two LGRB host galaxies not included in \cite{xshooter_survey}}
\end{minipage}
\begin{tabular}{m{1.5cm}L{1.2cm}L{1.8cm}L{1.8cm}L{1.8cm}L{1.8cm}L{1.8cm}L{1.8cm}L{1.8cm}L{1.8cm}}
\hline
\hline
\multirow{2}*{GRB host} & \multirow{2}*{Redshift} & \multicolumn{2}{c}{[\ion{O}{2}]} & H$\beta$ &  \multicolumn{2}{c}{[\ion{O}{3}]} & H$\alpha$ &  [\ion{N}{2}]  \\
\cline{3-4}
\cline{6-7}
           &            & $\lambda$3726 & $\lambda$3729 & $\lambda$4861 & $\lambda$4959 & $\lambda$5007 & $\lambda$6563 & $\lambda$6584 \\
\hline
GRB000210 & 0.8456 & $2.5 \pm 0.3 \pm 0.3$& $4.2 \pm 0.4 \pm 0.5$ & $1.6 \pm 0.1 \pm 0.2$ & $1.4 \pm 0.4 \pm 0.2$& $4.8 \pm 0.3 \pm 0.6$ & $6.8\pm 0.5\pm 0.9$ & $1.1 \pm 0.9 \pm 0.1$\\
GRB011211$^{\rm{a}}$ & 2.1435 & $1.1 \pm 0.5 \pm 0.3$ &  $1.4 \pm 0.1 \pm 0.4$ & $0.7 \pm 0.6 \pm 0.2$ & $1.6 \pm 0.3 \pm 0.4$& $5.9 \pm 1.2 \pm 1.5$& $3.5 \pm 1.3 \pm 0.9$ & $1.0 \pm 1.1 \pm 0.3$\\ 
\hline
\end{tabular}
\end{center}
\vspace{-0.3 cm}
\begin{minipage}[H]{1\textwidth}
{\small {\bf Notes.} Measurements are in units of 10$^{-17}$~erg~cm$^{-2}$~s$^{-1}$ , and are corrected for the Galactic foreground reddening, but not for intrinsic host galaxy extinction. Corrections for slit-loss based on broad-band photometry have been applied to the measurements following the same method as in \cite{xshooter_survey}. The first error represents the statistical error due to photon statistics and line-flux measurement. The second error is the systematic error in the absolute flux calibration due to slit-loss and scaling to photometry.
$^{a}$ The individual components of the [\ion{O}{2}] doublet are poorly constrained, either because of telluric absorption, skylines, or low signal-to-noise level.}
\end{minipage}
\end{table*}

The X-Shooter integrated IR spectroscopic channel extends the redshift coverage of the LGRB host metallicity sample out to {\it z} = 2.47\footnote{This is discounting higher redshift objects at $z>2.5$ in \cite{xshooter_survey} due to use of the [Ne III] method for breaking R$_{23}$ metallicity degeneracy which should never be applied to LGRB hosts \citep{stats_paper}.}. Thus, in combination, \cite{xshooter_survey} and \cite{stats_paper} provide a sample of sufficient size and redshift range to allow division into { 5} different redshift bins{, spanning $0$~$<$~$z$~$<$~$2.5$, allowing for detailed} study { on} the evolution of the LGRB { host} metallicity distribution with redshift as we will do here.

\subsection{Redshift distribution}\label{Host_redshifts}
\begin{figure*}[t]
\begin{center}
\includegraphics[width=.85\textwidth]{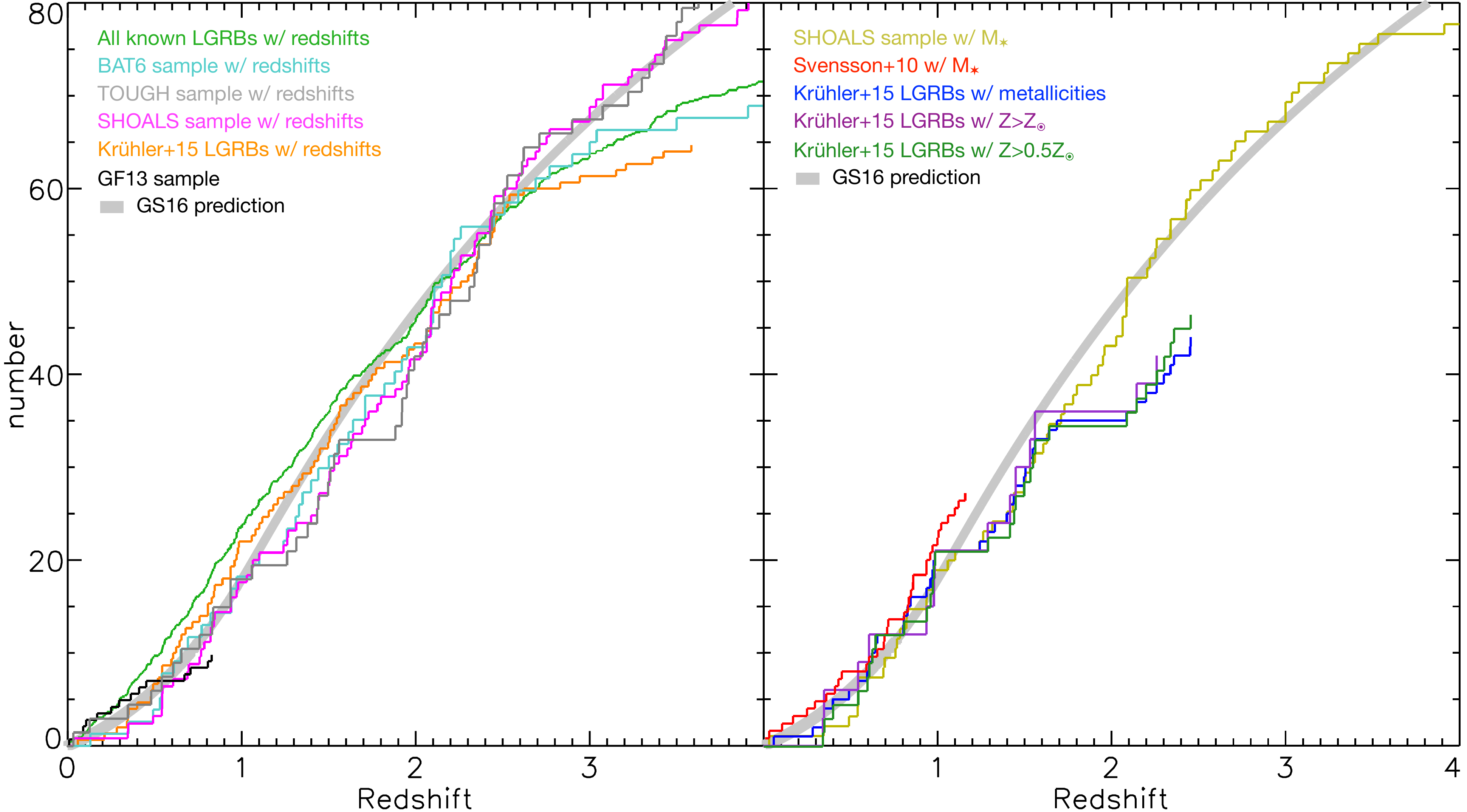}
\caption{\label{redshift_distribution} The redshift distribution of various LGRB samples normalized such that the distributions have a value of approximately 60 at a redshift of 2.5. The scaling factors applied are given in \aref{redshift_scaling} of \aref{AppendixA}. Deviations between the sample distributions allow selection effects within the samples to be identified. A gap is seen in the X-shooter metallicity sample (both the total sample, blue line, and its high metallicity sub-samples, purple and sark green lines) from 1.7~$<$~{\it z}~$<$~2.1 due to observational difficulties in obtaining all the lines needed for metallicity measurement (see text).  As measuring the redshift does not require observing specific lines this gap is not seen in the \cite{xshooter_survey} redshift sample. Note that the \cite{xshooter_survey} super solar metallicity LGRB sample (purple) follows the same distribution as the general \cite{xshooter_survey} metallicity sample (blue), which suggests both that (i) the metallicity of the LGRB hosts does not bias the likelihood of being able to measure this metallicity (i.e.\ low metallicity galaxies have metallicity measurements at the same rate as high metallicity ones), and (ii) the distribution of metallicities measured in LGRB host galaxies does not notably evolve out to $z \sim$ 2, which is consistent with the results of \aref{mzd}.}
\end{center}
\end{figure*}

Our LGRB host galaxy metallicity sample covers the redshift range $0.009<z<2.454$. The redshift cumulative distribution is shown in \aref{redshift_distribution} together with several other LGRB host galaxy samples, where we have scaled each population by a unique arbitrary factor to line up with the \cite{xshooter_survey} metallicity distribution, thus producing distributions consistent with each other. The name or reference for each sample is indicated in the figure legend, followed, where relevant, by the property used to select a subsample. For example, the label `Kr{\"u}hler+15 LGRBs w/ redshifts' refers to the sample of LGRB host galaxies taken from \cite{xshooter_survey} that have available redshifts.\footnote{\cite{xshooter_survey} excludes 9 LGRB hosts due to their having no detected stellar continuum or emission-lines precluding determining an emission line host redshift.  Two of these 9 objects LGRBs 090926A and 101219B do have detected afterglow redshifts (of $z$ = 2.11 and 0.55 respectively) but they have not been added back into the `Kr{\"u}hler+15 LGRBs w/ redshifts' sample.} Further details on each of the comparison samples are given in \aref{AppendixA}. Of interest in this figure are the differences in the shape of the distributions between the various populations shown. A difference in the slope between samples indicates a redshift?dependent bias between the populations, in particular a deviation in the slope of a single sample would indicate the presence of bias within that sample in the corresponding redshift range.  Thus by comparing populations of galaxies with different observed properties, we can identify any clear redshift biases present in any of the samples considered here.

\begin{figure}[h]
\begin{center}
\hspace{-0.03\textwidth}\includegraphics[width=0.5\textwidth]{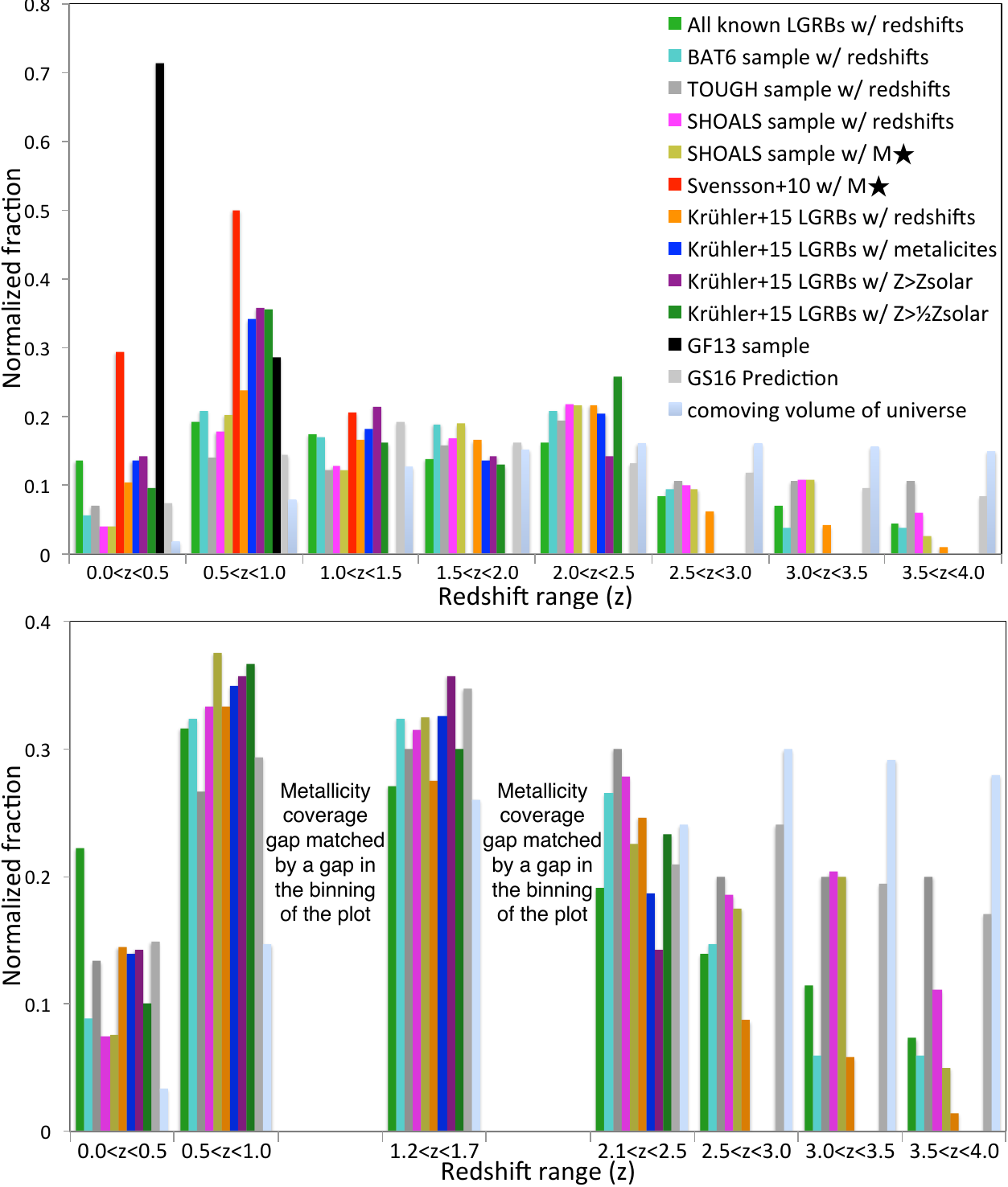}
\caption{\label{bar_hist_final} Normalized number of objects of each type in \aref{redshift_distribution}.  The upper chart is normalized to the total number of ({\it z} $<$ 4) objects  and uses regular binning.  The lower chart is normalized to the total number of {\it z}~$<$~2.5 objects (dropping the 2 samples which don't approach this redshift) and the binning has been adjusted around the redshift metallicity gaps.  The comoving volume of universe has been added for reference.  Legend applies to the lower chart as well and the colors of the populations are matched to \aref{redshift_distribution}.}
\end{center}
\end{figure}

\begin{figure*}
\begin{center}
\includegraphics[width=0.8\textwidth]{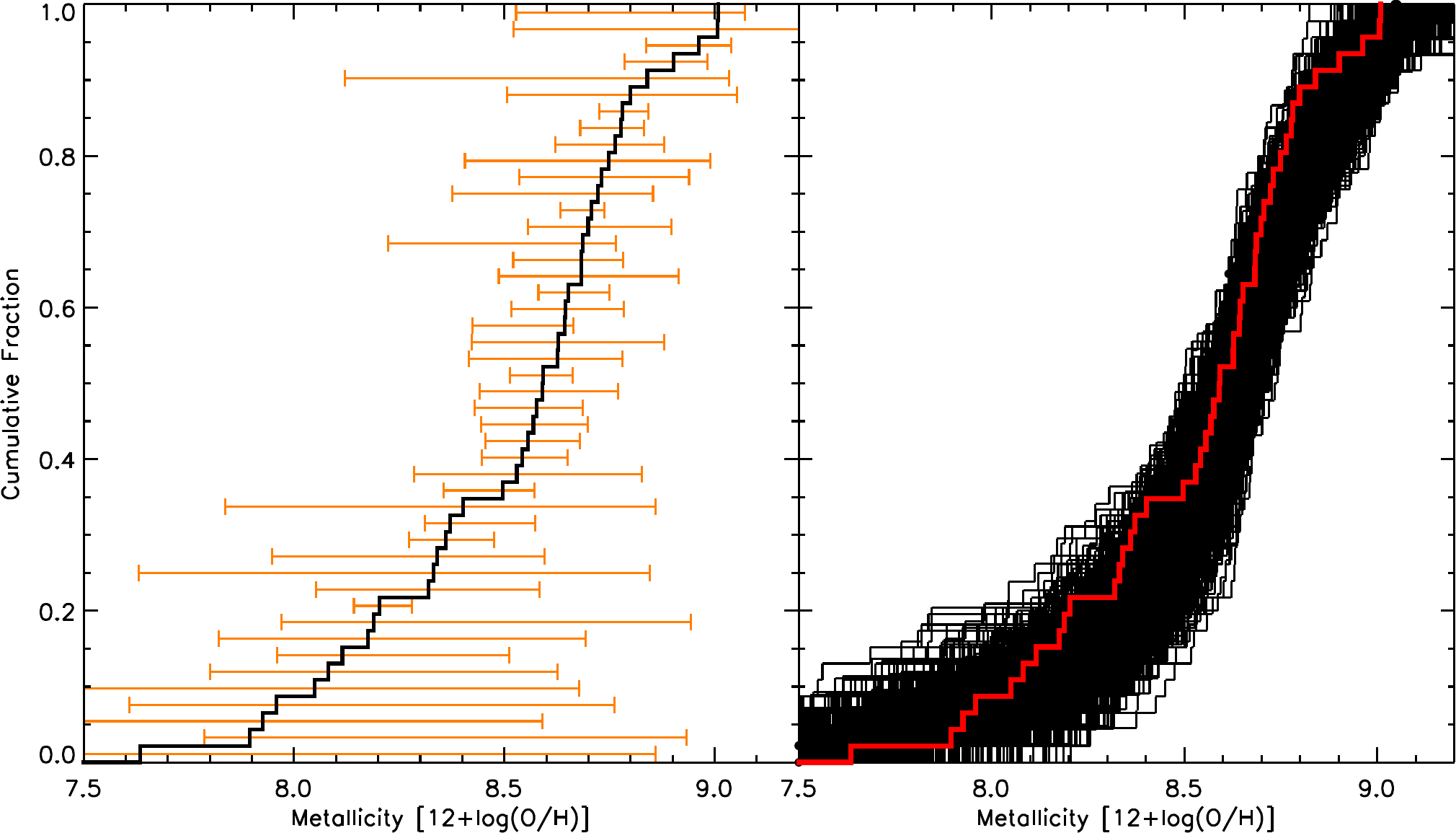}
\caption{\label{clum_metallicity_w_error_sbs} {\it Left}: cumulative distribution of objects for which we have measured metallicities with errors (see \aref{xsd_data_table}) with the error range of each value overplotted.  {\it Right}: The 1000 metallicity cumulative distribution for the 55 LGRBs in our sample produced by the 1000 Monte Carlo iterations described in \aref{Metallicity_Errors}. For reference, the black line in the left plot is overplotted in red in the right plot.  Note that both the errors for the individual objects and the cumulative distribution errors are not symmetric.}
\end{center}
\end{figure*}

Overall there is surprisingly good agreement between the redshift distributions shown in \aref{redshift_distribution}. This can also be seen in \aref{bar_hist_final}, where we plot the same redshift samples, but now in bar chart form. However, there are some features of note. Namely, the \cite{xshooter_survey} metallicity sample has a gap from 1.7 $\lesssim$ {\it z} $\lesssim$ 2.1 which is due to a confluence of observational limitations.  The [\ion{O}{2}]$\lambda\lambda$3726,2729 line doublet is redshifted to the optical IR transition at about 1 micron, and the H$\beta$ and [\ion{O}{3}$\lambda\lambda$4959,5007 lines are all redshifted to a region of poor sky transmittance between $J$ and $H$ bands. The H$\alpha$ line is redshifted into a similar low transmittance region between the $H$ and $K$ bands.  We also find a smaller gap from 1.0 $\lesssim$ {\it z} $\lesssim$ 1.2 due to similar effects.

The samples diverge at {\it z} $\gtrsim$ 2.5 due to different completeness rates for high redshift objects.  As this is beyond the redshift range of our host metallicity sample, it does not impact our subsequently metallicity distribution analysis.  It is worth noting that the \cite{Perley_shoals_masses} SHOALS and the TOUGH samples do not show a change of slope in these regions, unlike that seen in the BAT6 sample and in the sample of all LGRBs with known redshifts, and instead agrees quite closely with our predicted LGRB event rate curve from the \cite{form_rate_letter} formulation. Since the BAT6 sample is flux limited, as they go higher in redshift, their sensitivity to the lower end of the GRB luminosity function decreases.

\subsection{Host Metallicities}\label{Host_Metallicities}
Due to differences between metallicity diagnostics and their various calibrations, a comparison of metallicity values requires that they be determined using as consistent a method as possible. Therefore we recalculate the \cite{xshooter_survey} metallicities, from their published flux values, using the R$_{23}$ metallicity diagnostics in the \cite{KobulnickyKewley} (KK04) scale, using the exact same code as used to determine the metallicities given in \cite{stats_paper}.

The R$_{23}$ method is one of the primary metallicity diagnostics for galaxies at $z>0.3$.  The R$_{23}$ diagnostic is based on the electron temperature sensitivity of the oxygen spectral lines, and uses the ratio of the oxygen line strength to a spectral feature independent of metallicity.  R$_{23}$ requires measurement of the [\ion{O}{2}]$\lambda\lambda$3726,3729, [\ion{O}{3}]$\lambda$4959, [\ion{O}{3}]$\lambda$5007, and H$\beta$ lines.\label{metal_lines}  However as the Oxygen line strength initially increases with, and then decreases with increasing metallicity (due to infrared fine-structure lines inducing a cooling effect), this method alone is only sufficient to determine a pair of degenerate upper and lower branch metallicity values.  Therefore the R$_{23}$ diagnostic is typically coupled with a second metallicity diagnostic to resolve this degeneracy. Typically, and specifically for all the objects in this paper, this is done using the [\ion{N}{2}]/H$\alpha$ diagnostic requiring spectral coverage of the [\ion{N}{2}]$\lambda$6584 to H$\alpha$ line ratio (or its limit).  While the [\ion{N}{2}]/H$\alpha$ diagnostic is not nearly as precise as R$_{23}$, it is sufficient to exclude one of the R$_{23}$ branches except when the branches are near their intersection point, where the metallicity values of the two branches converge anyway (the metallicity of the branch convergence point is dependent on the ionization but is typically around 12~+~log(O/H) $\approx$ 8.4, see \citealt{KobulnickyKewley}). The measured metallicity values used in this work are given in \aref{xsd_data_table}.

\subsubsection{Metallicity Errors}
\label{Metallicity_Errors}

Simple error propagation cannot be applied to the \cite{KobulnickyKewley} (\citetalias{KobulnickyKewley}) metallicity diagnostic because it involves an iterative process (unlike metallicity diagnostics which use a formula of the line strengths such as done with the \citealt{Dopita2016}.  The iterative approach is necessitated because the relation of the line fluxes to metallicity is dependent on the ionization parameter whose relation to the line fluxes is dependent on the metallicity, therefore the code begins with an initial metallicity estimate and then iteratively refines it to the output values.

We derive metallicity errors by running Monte Carlo iterations with error perturbations of all lines simultaneously, where the perturbations are generated at random from a normal distribution with width equal to the corresponding line flux uncertainty, $\sigma$. Specifically, for each spectral line we independently calculate the inverse standard error function of a uniformly distributed random number, $u$, between 0 and 1 (erf$^{-1}(u)$), and multiply this by a factor of $\sigma\sqrt{2}$. This then gives an uncertainty value that has a corresponding probability $u$ of being generated from a normally distributed function with standard deviation $\sigma$. We then add this uncertainty, $\sigma\sqrt{2}$~erf$^{-1}(u)$, to our measured line flux, and use this to calculate a new \citetalias{KobulnickyKewley} metallicity for each individual Monte Carlo iteration. The difference between the original, and perturbed metallicities are then sorted and the values containing the 68.27\% of the sample closest to zero are adopted as the (positive and negative) error value for the object's metallicity.

It should be noted that the errors on our metallicities are not symmetric, and the reason for this is that the metallicity branch uncertainty causes some objects to have wildly asymmetric errors that reach out to cover the other metallicity branch. This is primarily driven by the uncertainty in the [\ion{N}{2}]$\lambda$6584 line flux and, as the [\ion{N}{2}]/H$\alpha$ ratio is a function of metallicity, this tends to result in larger errors for low metallicity objects. Thus objects, especially those with large line flux uncertainties, tend to have the derived metallicity uncertainties biased in the direction of the typical metallicity range (i.e.\ low metallicity objects have errors biased high and high metallicity objects have errors biased low). Our method of determining the uncertainties is similar to what has been done in other papers that use more complex metallicity diagnostics where it is not possible to apply simple error propagation \cite[e.g.,][]{new_metal_code}. For example, methods that aim to simultaneously minimise the difference between the observed and predicted line ratios using a number of metallicity diagnostics \citep[e.g.,][]{Nagao,Maiolino2008,Blanc2015,Curti2020} will often use Markov chain Monte Carlo methods to sample the probability distribution function of each set of relevant line ratios or corresponding metallicity values  \citep[e.g.,][]{xshooter_survey,Mingozzi2020,Curti2020}.

The metallicity errors calculated for each of the objects in our sample are plotted on top of the cumulative distribution of the metallicities in \aref{clum_metallicity_w_error_sbs} left. In \aref{clum_metallicity_w_error_sbs} right we plot the 1000 metallicity cumulative distributions generated from each of our Monte Carlo runs, where the cumulative distribution of the `unperturbed' metallicities is shown in red for comparison.

\section{Metallicity Distribution as a Function of Redshift}
\label{mzd}

\begin{figure*}[t]
\begin{center}
\begin{minipage}[t]{0.45\textwidth}
\includegraphics[width=1\textwidth,height=1\textwidth]{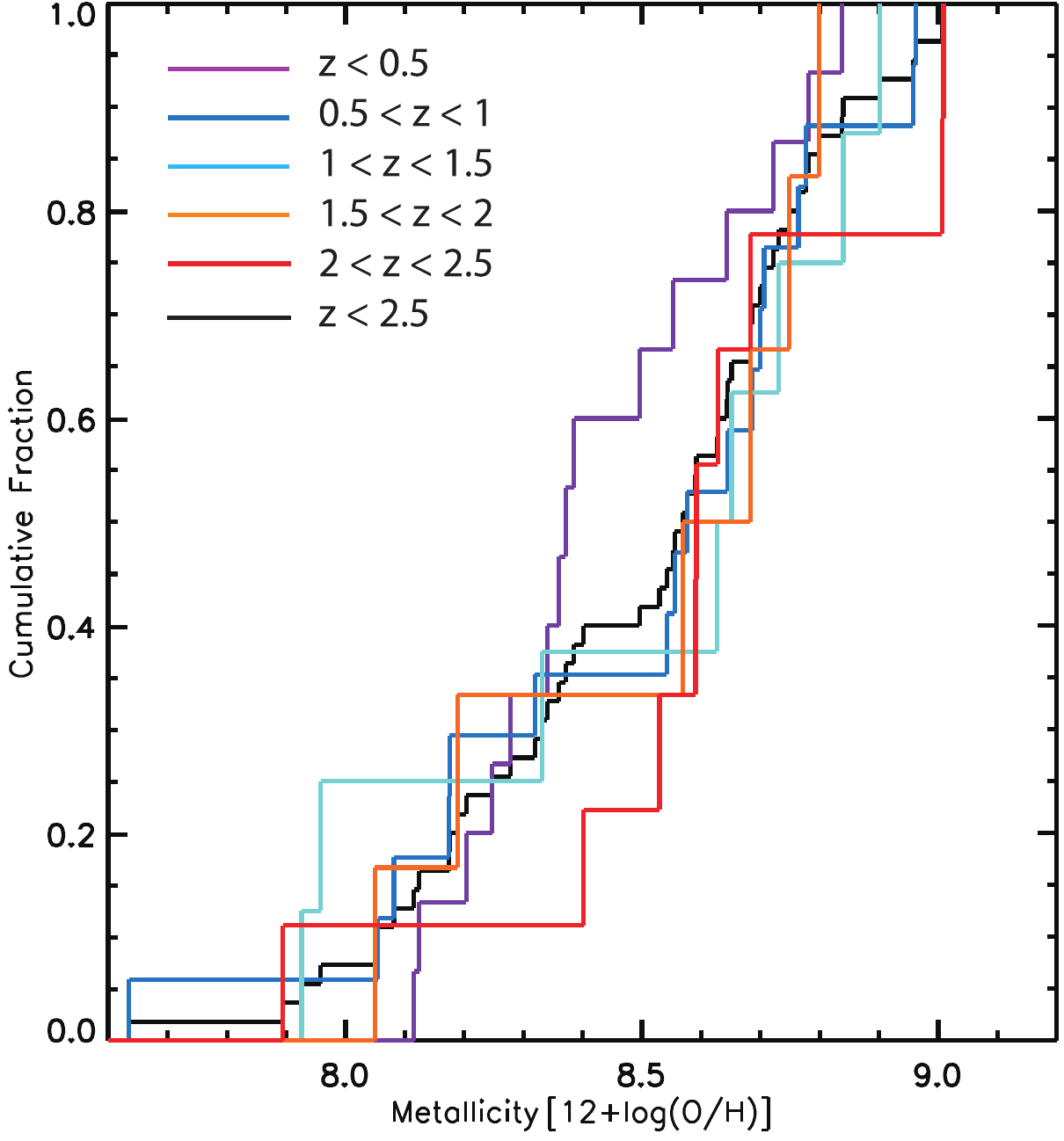}
\caption{\label{measured} Cumulative distribution of measured LGRB host metallicities binned by redshift. The sample is selected from \cite{stats_paper} and \cite{xshooter_survey} (see \aref{sample}), and all metallicities have been computed using the R$_{23}$ diagnostic and \citealt{KobulnickyKewley} (KK04) scale and code.  Surprisingly, the data show no evidence for statistically significant evolution in the LGRB host metallicity distribution with redshift.}
\end{minipage}
\begin{minipage}[t]{0.03\textwidth}
~
\end{minipage}
\begin{minipage}[t]{0.45\textwidth}
\includegraphics[width=1\textwidth,height=1\textwidth]{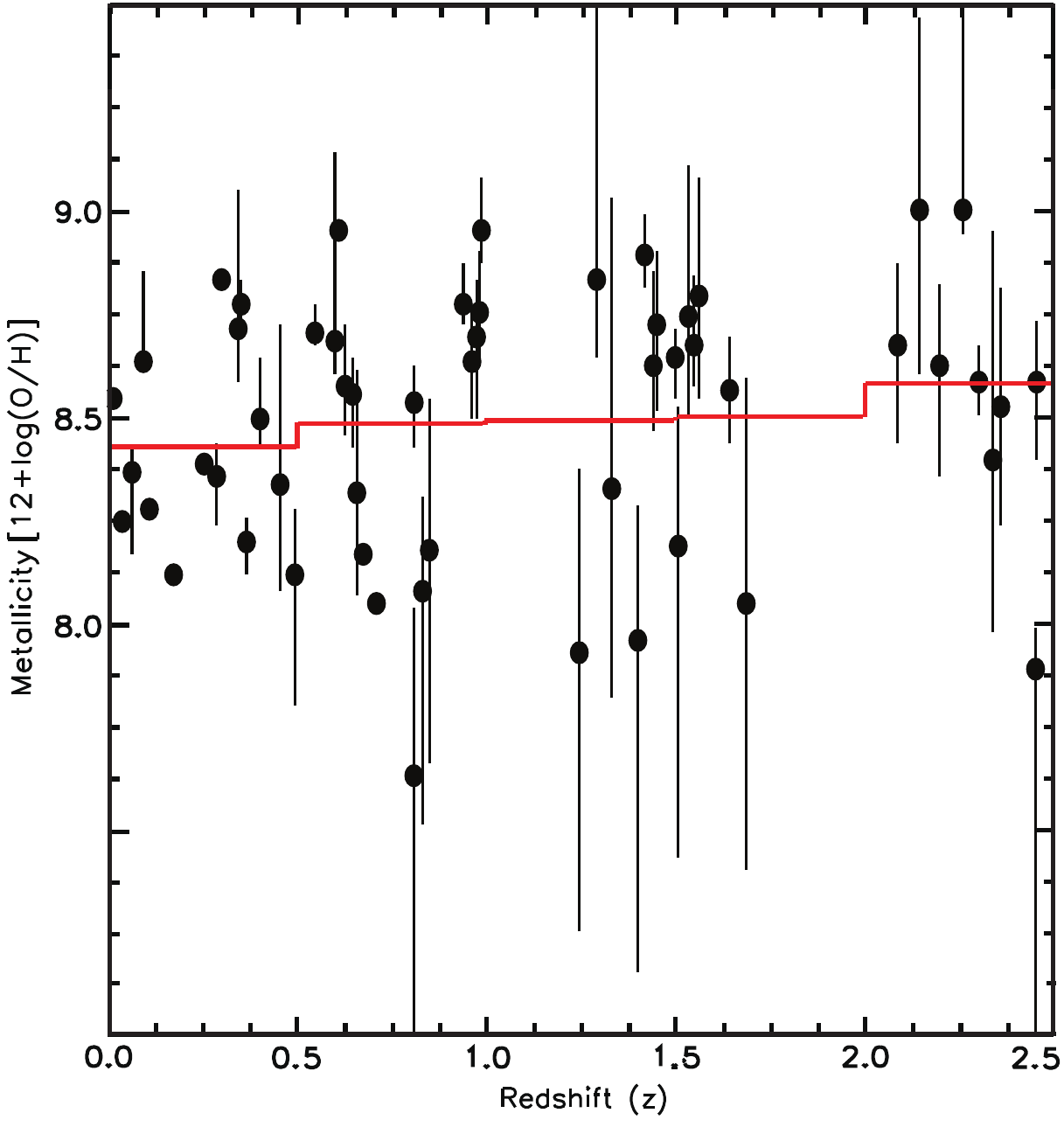}
\caption{\label{metallicity_vs_redshift} Metallicity as a function of redshift for the 55 LGRB host galaxies in our sample. The oxygen abundance is computed using the \citetalias{KobulnickyKewley} R$_{23}$ diagnostic described in \aref{Host_Metallicities}. The average R$_{23}$ metallicity in bins of $\Delta z =0.5$ is plotted in red and given in \aref{hist_mean_table} to illustrate the lack of evolution in metallicity with redshift.}
\end{minipage}
\end{center}
\end{figure*}

\begin{figure*}[th]
\begin{center}
\includegraphics[width=1\textwidth]{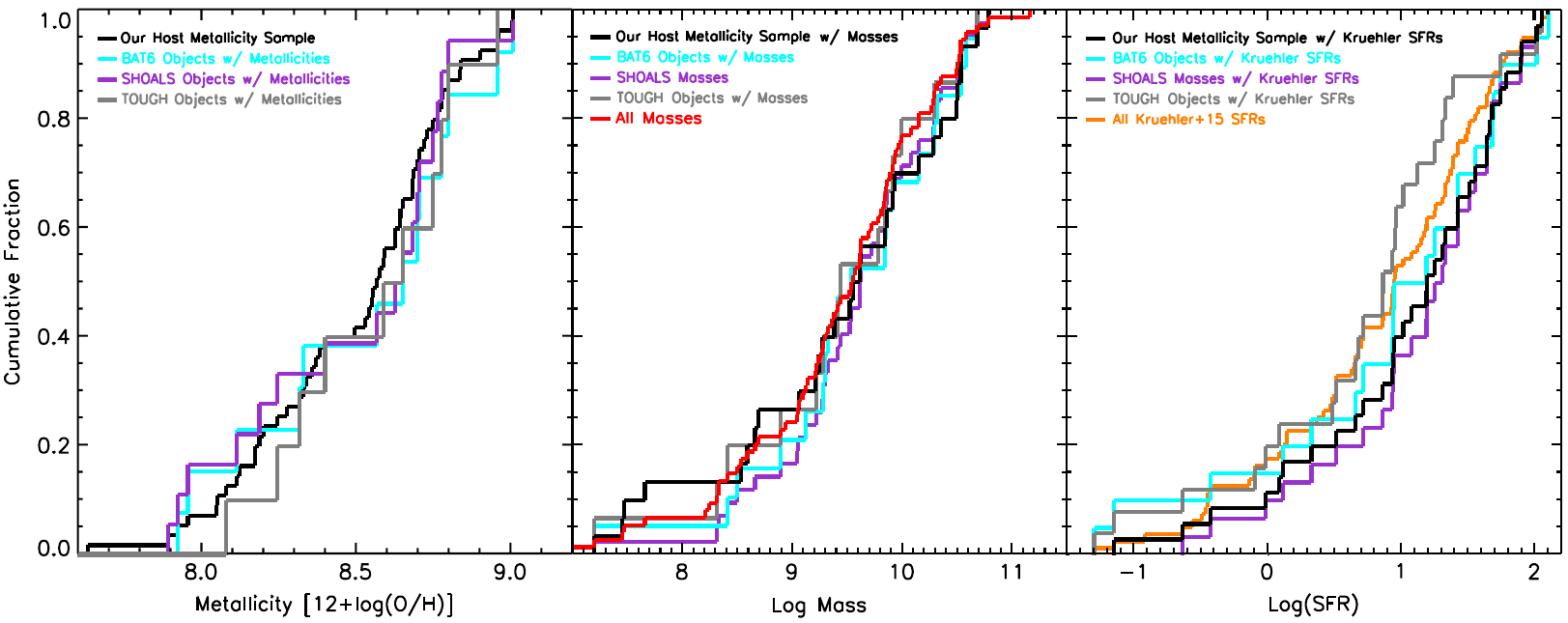}
\caption{\label{sample_comp_comb_plot} Cumulative distribution plots comparing the host metallicity (left), mass (center), and SFR (right) of the LGRB host metallicity sample compiled here with the subset of LGRB host galaxies taken from the \cite{BAT6} BAT6, \cite{Perley_shoals_masses} SHOALS, and \cite{TOUGH} TOUGH samples. All samples are limited to the same redshift range as our LGRB host galaxy sample ($z<1.7$ and $2.1<z<2.5$). We find quite good agreement between the metallicity and mass distributions of the various samples, and reasonable agreement in SFR. See \aref{Selection_Effects} for a more detailed discussion of the comparisons used and their results. K-S test results for the left, center, and right plots are given in \aref{metallicity_sample_comp_KS_table}.}
\end{center}
\end{figure*}

In order to track how the metallicity distribution of LGRB hosts varies with redshift, we compute the normalized metallicity cumulative distribution of our LGRB host galaxy sample at different redshift bins. This is only possible due to the large size of our LGRB sample. In \aref{measured} we plot the resulting cumulative metallicity distribution plots divided across five redshift bins with a width of $\Delta z=0.5$. The {\it z}~$>$~2.0 sample appears to be a little more metal rich than the 0.5~$<$~{\it z}~$<$~2.0 sample, but this is not statistically significant, with the greatest difference being between the $z<0.5$ and the $2<z<2.5$ samples, which, based on a Kolmogorov-Smirnov (K-S) test, have a 10\% probability of stemming from the same parent population. Furthermore, any increase in the metallicity of LGRB host galaxies with redshift is counter intuitive, given the typically lower metallicities of high-z, star-forming galaxies. This lack of metallicity evolution can also be seen in the plot of metallicity against redshift shown in \aref{metallicity_vs_redshift}.

The overall result is a markedly constant metallicity distribution across the different redshift bins, indicative of far less metallicity evolution than is present in the typical star-forming galaxy population across the same redshift range \citep{Zahid2013}. Since LGRBs are formed in star-forming galaxies this discrepancy is perplexing.

To verify the dependency of this result on our choice of metallicity diagnostic, we repeated this analysis but now measuring the galaxy metallicities using the relatively new empirical calibrations for six gas-phase metallicity diagnostics from \cite{Curti2017} (\citetalias{Curti2017}). All six diagnostics were calibrated against the $T_e$-based metallicities measured in the same sample of $\sim 120,000$ galaxies taken from the Sloan Digital Sky Survey (SDSS) data release 7 \cite[DR7;][]{SDSS_DR7}. Instead of providing six metallicities per LGRB host galaxy, we used the method described in \cite{Curti2020} to determine the metallicity that provides the simultaneously best agreement to each of the six \citetalias{Curti2017} metallicity diagnostics for a given set of measured line fluxes and uncertainties. More details on the \citetalias{Curti2017} metallicity calibrations and \cite{Curti2020} method are given in \aref{AppendixB}. Using this alternative metallicity diagnostic, we find that the \citetalias{Curti2017}-based metallicities also show a lack of evolution with redshift in the LGRB host galaxy metallicity cumulative distributions (see \aref{BF_metallicity_CDFs}, left panel), consistent with our \citetalias{KobulnickyKewley} R$_{23}$ results, despite the two diagnostics having been calibrated in two very different ways.  To illustrate this lack of metallicity evolution in another way, in \aref{metallicity_vs_redshift} we also plot our measured LGRB host galaxy \citetalias{KobulnickyKewley} R$_{23}$ metallicities against redshift. In \aref{BF_metallicity_CDFs} we show similar plots but including the \citetalias{Curti2017} best-fit metallicities against redshift.

It is important to consider whether any selection effects present in our sample may be effecting the results shown in \aref{measured}. In particular luminosity biases, i.e.\ selecting only galaxies bright enough for spectroscopy in a reasonable amount of telescope time, would presumably be prevalent.  However the ratio of LGRB host galaxies observed at low, intermediate, and super solar metallicities seems to remain relatively consistent, particularly at {\it z} $<$ 2, (see \aref{measured}) whereas the general expectation for luminosity biases would be to observe an increasingly large fraction of higher metallicity galaxies with redshift. We cannot rule out a scenario where the lower metallicity LGRB hosts are being selected out via a luminosity bias, while the higher metallicity LGRB hosts become intrinsically less common with redshift, thus leading both populations to drop off at the same rate and maintain a constant ratio. However, we show in \aref{Z_frac} that there is little evidence for a drop in the population of higher metallicity LGRB host galaxies with redshift, making this explanation unlikely.

Ideally we would compare the metallicity distribution of LGRB hosts with a sample of typical star-forming galaxies at the same redshifts, selected in a star-formation weighted manner (i.e.\ the sample methodology of the  \cite{stats_paper} SDSS population but at higher redshifts).  Unfortunately, a suitable galaxy sample does not exist.  The SDSS metallicity sample only extends (with a large sample size) out to {\it z}~$<$~0.3 and even out to that redshift range there are completeness issues with faint galaxies. Therefore our ability to quantify the expected metallicity distribution evolution for the LGRB hosts from the typical star-forming galaxy population is lacking.

\subsection{Comparison with other LGRB host galaxy samples}
\label{Selection_Effects}
Given the inhomogeneity of our sample, we compare some key properties of our LGRB host galaxy sample to some other, carefully selected host galaxy samples in order to explore the presence of biases in metallicity, mass, and SFR.  While we expect most potential sample biases to distort the distribution of multiple physical parameters, including redshift, the additional cross-checks of these other physical properties preclude some possible biases.  In \aref{sample_comp_comb_plot} we plot the cumulative distributions in host metallicity (left), mass (center), and SFR (right) of our LGRB host metallicity sample and the presumably less biased \cite{BAT6} BAT6, \cite{Perley_shoals_masses} SHOALS, and \cite{TOUGH} TOUGH comparison samples. 

To maintain consistency in our comparisons, we used the metallicity measurements computed in this paper, the mass measurements from \cite{Perley_shoals_masses} and \cite{Svensson}, and the SFRs from \cite{xshooter_survey}. As the distribution of LGRB host masses varies with redshift (see \citealt{Perley_shoals_masses}), all comparison samples have been limited to the redshift range of the LGRB host metallicity sample ({\it z}~$<$~1.7 or 2.1~$<$~{\it z}~$<$~2.5).  We have deliberately limited this analysis to using a single sample of mass and SFR values for the same reason we recalculate all the metallicity values used in this paper directly from line fluxes; so as to preclude introducing differences in methodology. We select the \cite{xshooter_survey} SFRs and \citealt{Perley_shoals_masses} masses as they are the largest such samples. Although biases may have been introduced by only selecting GRB hosts from the BAT6, TOUGH and SHOALS samples with the reported galaxy properties of interest, the selection functions of the parent population are at least well understood. In \aref{comp_comb_plot_num_table}, we provide the total number of objects in our sample and the comparison samples within the redshift range $z<1.7$ and $2.1<z<2.5$ considered here. We also provide the size of each of the subsamples that have each of the corresponding measured parameters shown in each panel of \aref{sample_comp_comb_plot}, and the associated sample completeness. 

In the metallicity (left most) plot, we plot all the LGRB host metallicities in our sample (\aref{xsd_data_table}) together with the subset of those LGRB host galaxies in the BAT6, SHOALS, and TOUGH samples that have metallicities (and within the same redshift range as our sample). In this comparison, we use nearly all published host galaxy spectroscopic data from which a \citetalias{KobulnickyKewley} R$_{23}$ metallicity can be computed. To our knowledge, the only exception is a single object from \citealt{Palmerio2019} (GRB~061121). Any remaining biases in our sample would thus be present in all studies that investigate the distribution of LGRB host galaxy metallicities.

\begin{table}
\begin{center}
\vspace{-0.2 cm}
\begin{minipage}[H]{0.47\textwidth}
\caption{\label{comp_comb_plot_num_table} Size (and completeness) of subsamples  in \aref{sample_comp_comb_plot}}
\end{minipage}
\begin{tabular}{lcccc}
\hline
\hline
 &  This Work & BAT6 & SHOALS & TOUGH \\
\hline
Metallicity &   55 (100\%) & 13 (38\%) & 18 (43\%) & 10 (32\%)\\
Log Mass &  30 (60\%) & 22 (65\%) & 42 (100\%) & 15 (48\%) \\
Log SFR &   35 (70\%) & 20 (59\%) & 30 (71\%) & 25 (81\%) \\
\hline
\end{tabular}
\vspace{-0.3 cm}
\begin{minipage}[H]{0.47\textwidth}
\small Number of LGRB host galaxies within the redshift range {\it z}~$<$~1.7 and $2.1<z<2.5$ in our sample, and the unbiased BAT6, SHOALS and TOUGH samples, and that have the property specified in the first column of the table. The completeness of each subsample within the redshift range of interest is given in parentheses.
\end{minipage}
\end{center}
\end{table}

In the mass (center) plot first, all the SHOALS masses (in the redshift range of the LGRB sample) are plotted, together with the subset of LGRB hosts in our sample, and in the BAT6 and TOUGH samples that have SHOALS masses. 

In the SFR (right most) plot, all the \cite{xshooter_survey} Table 4 LGRB host SFRs are plotted, along with the subset of LGRBs from our sample, and the BAT6 and TOUGH samples that have SFRs from \cite{xshooter_survey}. The right plot uses a simple {\it z}~$<$~2.5 redshift limit, however the results are consistent if the same redshift cut as the center plot is applied. The results are also unaffected if we consider only SFRs calculated from the H$\alpha$ line flux \citep[see Table 4 of][]{xshooter_survey}.

We find quite good agreement between the metallicity distributions of the various samples (see K-S results in \aref{metallicity_sample_comp_KS_table}), with K-S tests indicating that all distributions are consistent with being drawn from the same parent sample (see \aref{metallicity_sample_comp_KS_table}). The mass distributions are similarly consistent (all K-S tests indicate a p-value $>0.93$ of being drawn from the same parent population). Only the comparisons in the SFR show more significant differences in the cumulative distribution functions. However, even then, the two most divergent distributions (SHOALS and TOUGH) have a p-value p=0.1, meaning that we cannot rule out the null hypotheses that these two samples are drawn for the same parent population of galaxies. From these comparisons, there is therefore no clear, strong bias in the metallicity distribution of the LGRB sample considered here.

\subsection{Fraction of High Metallicity LGRB Hosts}
\label{Z_frac}
While the agreement between the cumulative distributions of metallicity at different redshift bins (\aref{mzd}), and between key host galaxy properties when compared to other LGRB galaxy samples (\aref{Selection_Effects}) is impressive, there are likely selection effects which we can not fully exclude. For example, the cutoff in absolute luminosity below which it is no longer practical to obtain a host galaxy spectrum will differ for each of the redshift bins. In this section we therefore consider the fraction of only the most metal-rich LGRB host galaxies, under the hypothesis that biases should dominate at low luminosities. Host galaxy metallicities are generally easier to obtain for more luminous and more metal-rich galaxies at all redshifts, and the redshift evolution in the fraction of LGRB hosts with known redshift that are metal-rich should therefore be less susceptible to selection effects.

\begin{table*}
\begin{center}
\begin{minipage}[H]{1\textwidth}
\caption{\label{calc_table} Table of Values Used In Computation of Ratios}
\end{minipage}
\begin{tabular}{L{0.3cm}m{8.0cm}L{2.2cm}L{2.2cm}L{2.2cm}L{1.8cm}L{1.8cm}L{2.2cm}}
\hline
\hline
line & &  0.0~$<$~$z$~$<$~1.0 &  1.2~$<$~$z$~$<$~1.7\footnote{\begin{minipage}[H]{1\textwidth} Irregular bin spacing to avoid redshifts gaps in measured metallicites\end{minipage}} & 2.1~$<$~$z$~$<$~2.5\\
\hline
1 & Number of All LGRB Hosts with Redshifts\footnote{\begin{minipage}[H]{1\textwidth}Errors on object numbers assumed to be square root of the number of objects.\end{minipage}} & 155  & 78 &  55  \\
2& LGRB Hosts with Measured Metallicites & 32  & 14  & 8 \\
3& LGRB Hosts with Metallicites 12~+~log(O/H)~$>$~8.4 & 17  & 9  & 7  \\
4& Measured Metallicites / All LGRB Hosts & 0.206 $\pm$ 0.040  & 0.179 $\pm$ 0.052  & 0.145 $\pm$ 0.055 \\
5& Metallicites 12~+~log(O/H)~$>$~8.4 / All LGRB Hosts &  0.110 $\pm$ 0.039  & 0.115 $\pm$ 0.041  & 0.127 $\pm$ 0.051 \\
6 & Dark Burst Adj. 12~+~log(O/H)~$>$~8.4 / All LGRB Hosts &  0.087 $\pm$ 0.031    & 0.081 $\pm$ 0.028   & 0.078 $\pm$ 0.031   \\
\hline
\end{tabular}
\end{center}
\vspace{-0.6 cm}
\end{table*}

In Line 4 of \aref{calc_table}, we give the fraction of { LGRB hosts} with redshifts which also have metallicity measurements in our sample.  As one would expect, the fraction appears to decline continuously with increasing redshift, from about one-fifth in the first redshift bin to about one-tenth in the last.  However, in Line 5 we give the fraction of all GRBs which have a measured metallicity in our sample of log(O/H)+12 $> 8.4 $.  Strikingly, this fraction does not decline with redshift. This result is also observed when we repeat the analysis using the \citetalias{Curti2017} metallicities that we measure when applying the \cite{Curti2020} method, where the fraction of LGRB hosts with log(O/H)+12 $> 8.4$ that we measure is $0.14\pm 0.03$, $0.14\pm 0.05$ and $0.13\pm 0.05$, for the three consecutive redshift bins listed in \aref{calc_table}.

An important point to recognize is that there is an overabundance of ``dusty'' GRBs (bursts with $A^{GRB}_V>1$) that make up the \cite{xshooter_survey} spectroscopic sample, as noted by the authors. Such a bias could have the effect of artificially increasing the fraction of dusty GRB host galaxies that we calculate. In order to verify that such a bias is not somehow artificially increasing the fraction of metal-rich LGRB host galaxies at high-z (thus producing a flat high-metallicity fraction), we attempt to correct for this effect by following a similar approach as done by \cite{xshooter_survey}. In that paper, \cite{xshooter_survey} compared their sample with more representative surveys and determined this overabundance as a function of redshift (typically a factor of $\sim$30\%), which they then used as a correction factor on their sample.

In line 6 of \aref{calc_table} we have recalculated the ratios in line 5 after reducing the number of bursts in our combined sample from \cite{xshooter_survey} with log(O/H)+12 $> 8.4 $ by the appropriate overabundance factor.  We have not adjusted the part of our sample from \cite{diff_rate_letter}, because, as discussed by those authors, that sample had a typical number of dusty bursts.  Although the ratios all go down somewhat, the primary result remains, the fraction of bursts with metallicities log(O/H)+12 $> 8.4 $ remains essentially constant independent of redshift, at $\sim 8$\%. Furthermore, we find that the overabundance of dusty GRB afterglows in our LGRB host galaxy metallicity sample is far smaller than is the case in the full, \cite{xshooter_survey} sample, with our third redshift bin ($2.1<z<2.5$) having no LGRBs with an afterglow dust extinction $A_V>1$~mag. The corresponding corrections that we apply should therefore be considered upper limits, and the true evolution in the fraction of high metallicity LGRB host galaxies is likely to be closer to what is given in line 5 of \aref{calc_table} rather than the fractions given in line 6.

While there are strong biases in the determination of which hosts have their metallicity measured, they do not play an important role in the relatively constant `high' metallicity fraction that we find with redshift. Given the correlation between galaxy luminosity and metallicity, to have obtained this result artificially \cite{stats_paper} and \cite{xshooter_survey} would have had to preferentially observe faint hosts at low-redshift, selecting out the low-redshift, high metallicity population of galaxies. However, this is the exact opposite of the actual selection effects.  \cite{stats_paper} went after any host that had a usable spectrum.  There is not a bias here towards faint (i.e.\ low metallicity) hosts, rather the opposite.  Similarly, \cite{xshooter_survey}, started with a sample composed of TOUGH, BAT6, GROND, SHOALS LGRBs, as well as other GRB afterglows where they had X-shooter observations of either the afterglow or host galaxy. They excluded only objects for which neither emission lines nor the stellar continuum was detected.  So again there was a bias towards bright hosts. This bias persists at all redshifts. Any bright host that could have been observed by either \cite{stats_paper} or \cite{xshooter_survey} were observed by them. What determined the completeness of the sample of bright objects was telescope time, weather and declination.

To quantize the degree to which our metallicity sample (and the metallicity measurement process used in general) is biased towards larger and brighter host galaxies, we divide the unbiased and highly complete (94\%) SHOALS sample of host galaxy stellar masses \citep{Perley_shoals_masses} between those objects with metallicities in this sample and those without. We calculate that LGRB hosts without metallicity measurements have an average log mass of $\sim$9.3 M$_{\sun}$, compared to an average of $\sim$9.8 M$_{\sun}$ for those with metallicity measurements, implying that biases present in LGRB host galaxy metallicity measurements are indeed low mass, low luminosity population, as expected.

\subsection{Metallicity distribution from mass-metallicity relation}\label{emzd}

In order to further explore potential selection effects, we fit a mass-metallicity-redshift relation and use this to estimate the metallicity distribution from a comparatively less biased sample of host galaxy masses.  For this fitting, we again use the \cite{Perley_shoals_masses} SHOALS sample of 110 GRBs hosts with known redshift and estimated stellar mass, supplemented by the sample of 34 GRB host stellar masses from  \cite{Svensson} to increase the number of LGRB hosts at $z$~$<$~1. Two objects overlap between the samples (LGRBs 060218 and 080319B) whose log masses differ by 0.24 and -0.43 between the \cite{Svensson} and \cite{Perley_shoals_masses} estimates respectively. In our combined sample, we remove duplication, using the \cite{Perley_shoals_masses} stellar mass. This is largely to optimize consistency as the \cite{Perley_shoals_masses} sample is larger.

\begin{figure}[t]
\begin{center}
\includegraphics[width=.49\textwidth]{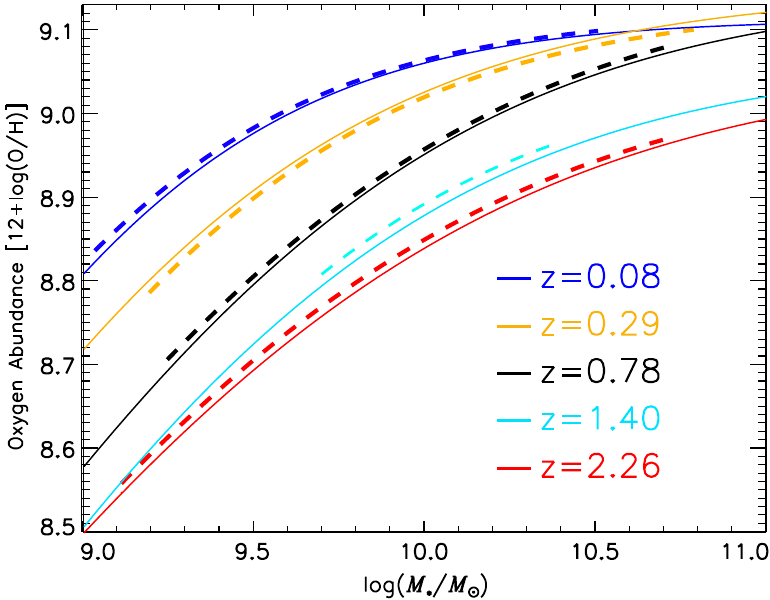}
\caption{\label{Zahid_polyfit} The mass-metallicity relation fits for different redshifts of \cite{Zahid2013} Figure 1 (thick dashed lines) with our estimated metallicity results for a continuous range of galaxy masses and matching redshift bins (thin continuous lines) overplotted.  The fitting is deliberately kept approximate to avoid over-fitting.}
\end{center}
\end{figure}

To estimate the MZR for the redshift of a given galaxy, we interpolate from the MZR measured at different redshifts. While \cite{Zahid2013} provide a series of mass metallicity relation fits across the redshift range of interest, we apply our own 2-dimensional fit to the data from \cite{Zahid2013} (see their Figure 1) in order to have a continuous model from which we can estimate the metallicity for a galaxy of any given mass and redshift. Care was taken to avoid over-fitting the \cite{Zahid2013} data and a number of fitting procedures were trialled with fitting a minimum curvature spline surface using the MIN\_CURVE\_SURF procedure adopted.  Our estimated metallicity results for a continuous range of galaxy masses are plotted against the individual redshift fits of \cite{Zahid2013} in \aref{Zahid_polyfit}. Although the metallicity distribution derived from the MZR reflects only the typical galaxy metallicity expected for a galaxy of a given LGRB host mass at its specific redshift, when applied to a sample of galaxies, as we do here, the intrinsic scatter in the relation should balance out.

\begin{figure*}[t]
\begin{center}
\begin{minipage}[t]{0.45\textwidth}
\includegraphics[width=1.05\textwidth]{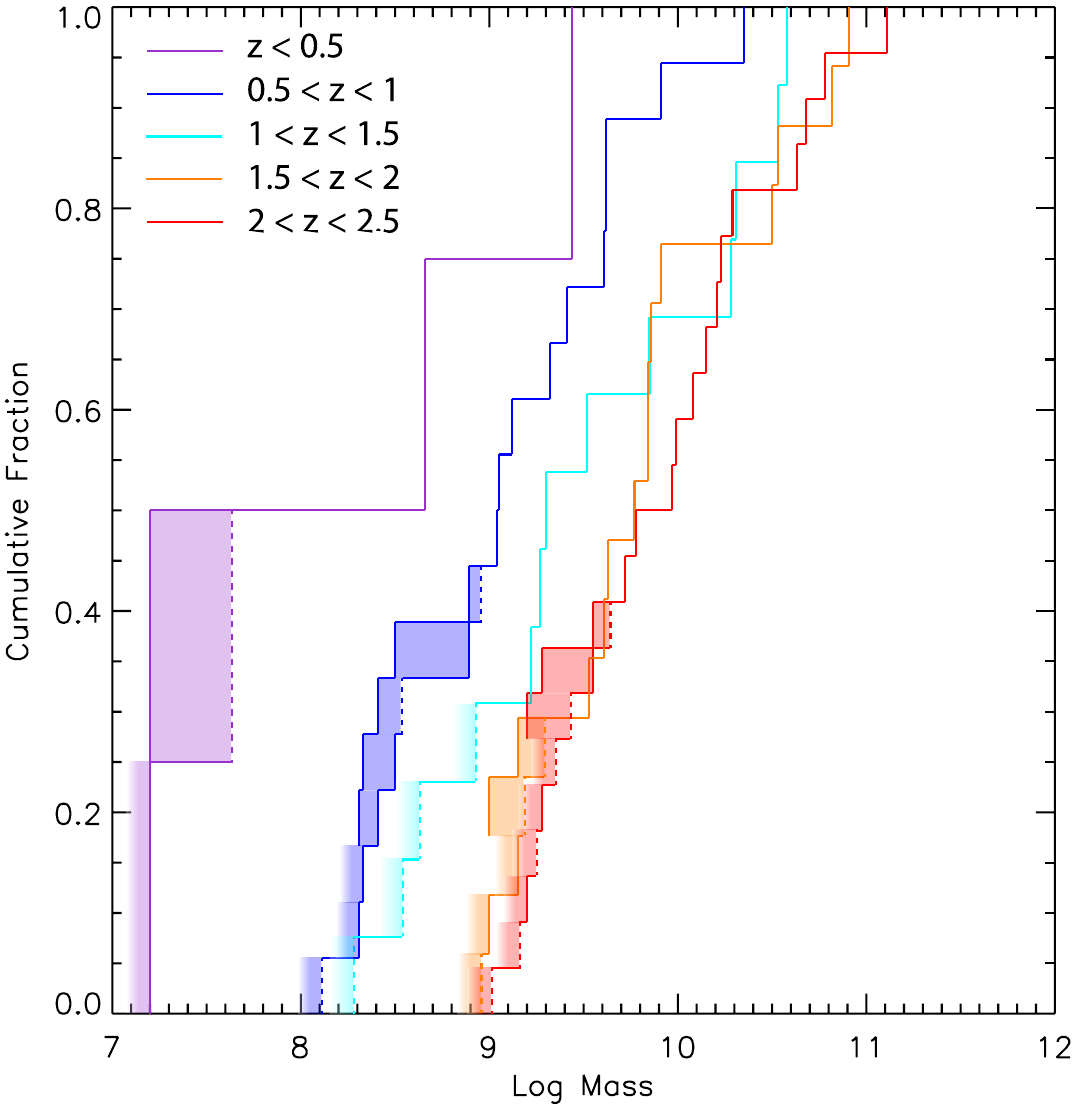}
\caption{\label{shoals_mass_step_plot_5}  Cumulative distribution of \cite{Perley_shoals_masses} SHOALS measured LGRB host masses binned by redshift. Mass upper limits are shown as dashed lines, mass measurements are shown as solid lines. When normalizing the cumulative distributions, we include the number of sources with just limits on their stellar mass, which is why the solid lines (corresponding to constrained stellar mass measurements) do not extend down to zero. These two approaches define the maximum and minimum range possible for the distributions.  A shaded region between them indicates where, due to the upper limits, the true distribution must lie.  Where the shaded region is bounded by lines (dotted or solid) the true distribution is constrained between those bounds.  If the color fades out to the left, the distribution could, in principle, extend to an arbitrarily low mass.
}

\end{minipage}
\begin{minipage}[t]{0.03\textwidth}
~
\end{minipage}
\begin{minipage}[t]{0.45\textwidth}
\includegraphics[width=1.05\textwidth]{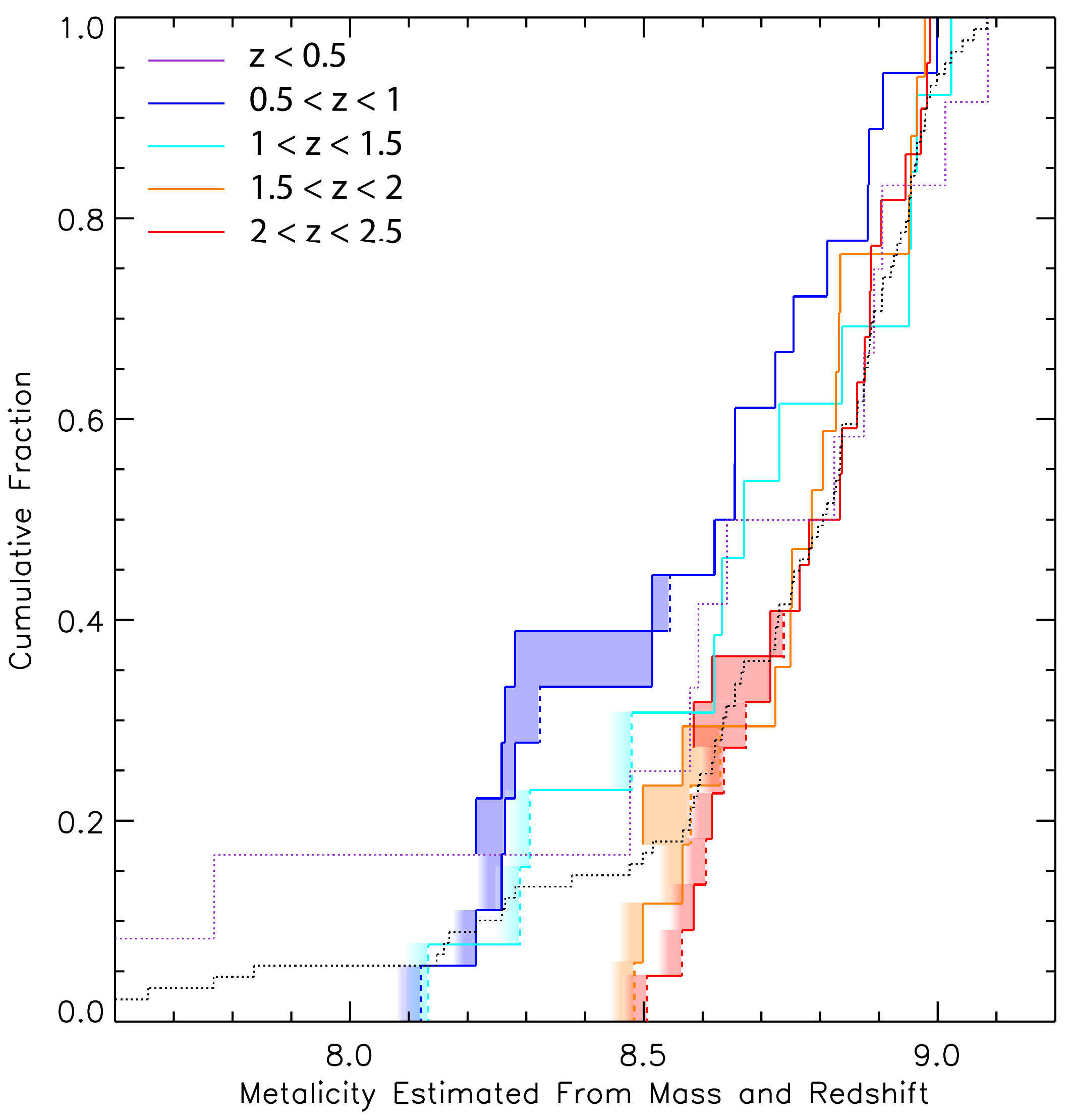}
\caption{\label{shoals_metal_est_step_plot_5}  Cumulative distribution of mass-based metallicity estimates from the \cite{Perley_shoals_masses} SHOALS mass and redshift values (see \aref{emzd}) binned by redshift. Estimated metallicity upper limits are shown as dashed lines, whereas estimated metallicity values (i.e.\ not limits) are shown as solid lines. As in \aref{shoals_mass_step_plot_5}, a shaded region indicates where the true distribution must lie. If the color fades out to the left, the distribution could, in principle, extend to an arbitrarily low metallicity.  Where the shaded region is bounded by lines (dotted or solid) the true distribution goes between the bounds. Due to the limited number of objects in the SHOALS $z$~$<$~0.5 bin we omit plotting this bin and instead replot the combined \cite{Perley_shoals_masses} and \cite{Svensson} $z$~$<$~0.5 bin.  For reference and comparison with full estimated metallicity sample, we also plot the combined metallicities across the entire $z< 2.5$ redshift range (dotted black line). Fortunately, all the upper limits are contained in the lower half of the distributions and thus do not disrupt this analysis.}
\end{minipage}
\end{center}
\end{figure*}

We apply this methodology to estimate the expected metallicities for the \cite{Perley_shoals_masses} SHOALS and \cite{Svensson} sample of LGRB host galaxy stellar masses. Although the SHOALS sample is produced using an unbiased selection criteria, seven out of 119 objects have either no redshift information (GRB~100305A) or only upper limits on their redshift. Furthermore, $\sim 25$ of the sample of objects with redshifts have only upper limits on their host galaxy mass values. We can address the latter shortfall by constraining the maximum and minimum possible distributions.

In \aref{shoals_mass_step_plot_5} we plot the stellar mass distribution of SHOALS galaxies with known redshift, including upper limits, using the same redshift binning as in \aref{measured}. In most cases, the SHOALS upper mass limits are below the majority of the measured mass values. We set the maximum possible distribution by plotting the data with measured mass values and mass upper limits together, and we indicate which values are limits by using a dashed line. The minimum possible distribution is set by assuming all the upper limit values are below the measured values. As there are no lower bounds on the upper limits, lower bounds are not plotted. However we do count the number of such objects and normalize the distributions accordingly. This results in distributions that do not extend all the way to zero (solid lines), but instead stop where the limits are unconstrained.  Where both sides are constrained, we shade the region between the maximum and minimum possible distributions, as the true distribution must lie between these bounds. When the minimum distribution is not constrained we use a gradient shading where the color fades out to the left as the distribution could, in principle, rise at an arbitrarily low mass. The resulting plot, \aref{shoals_mass_step_plot_5}, is similar to \cite{Perley_shoals_masses} Figure 4, but with added details to indicate the spread in the distributions due to the stellar mass upper limits, as described above. Of particular note in \aref{shoals_mass_step_plot_5} is that the host galaxy mass values continue to increase with redshift.

We use a similar technique when producing the cumulative distribution plots for the estimated metallicities, where upper mass limits now correspond to estimated metallicity upper limits, which are generally below the measured values. As can be seen from \aref{shoals_mass_step_plot_5}, the SHOALS-only $z$~$<$~0.5 bin has very few objects (just four), and we thus include stellar mass measurements for LGRB host galaxies at z$<0.5$ taken from \cite{Svensson} in order to increase the number of objective in this redshift bin. In \aref{shoals_mass_mzr_metallicities} we plot similar cumulative distribution plots of the LGRB host galaxy mass and the MZR-based metallicity estimates, but this time not including mass (and subsequent MZR-based metallicity) upper limits, which allows us to perform K-S tests on the various distributions. These are summarized in \aref{metal_est_comb_step_plot_5_KS_table} and \aref{metal_comp_table}.

The estimated metallicity cumulative distributions shown in \aref{shoals_metal_est_step_plot_5} are shifted increasingly to the right with redshift, corresponding to a constant or increasing metallicity distribution with redshift, in line with what we observe in \aref{measured}. Although an increase in observed host galaxy masses with redshift is expected due to Malmquist bias, the sample of LGRB hosts in \cite{Perley_shoals_masses} is unbiased in its selection, and 94\% complete in measured stellar mass. We note that the results shown in \aref{shoals_metal_est_step_plot_5} do depend on the MZR redshift evolution that we implement. There is disagreement in the literature on how quickly the MZR evolves with redshift \citep[e.g.,][]{Hunt2016, Yates2021}, with the evolution found by \cite{Zahid2013}, depicted in \aref{Zahid_polyfit}, being at the faster end \citep[e.g.\ see Figure~11 in][]{Yates2021}. If we were to implement a more slowly evolving MZR evolution in our analysis, such as was found by \cite{Hunt2016}, we would compute larger metallicities for the higher redshift galaxies in our sample, thus pushing the higher redshift bin cumulative distributions shown in \aref{shoals_metal_est_step_plot_5} further over to the right.

\begin{figure*}[t]
\begin{center}
\begin{minipage}[t]{0.45\textwidth}
\includegraphics[width=1\textwidth,height=1\textwidth]{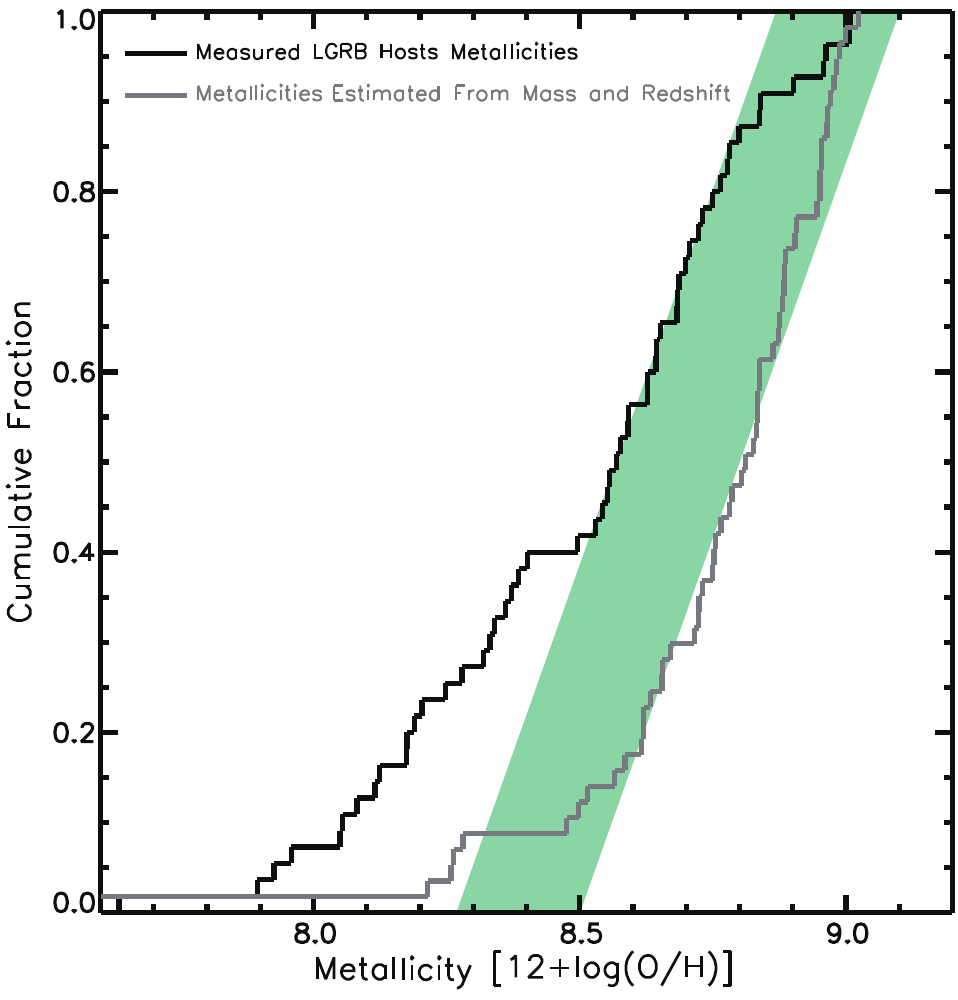}
    \caption{\label{measured_vs_estimated} Measured (black) and estimated (gray) metallicities from \aref{measured} and \aref{shoals_metal_est_step_plot_5}, respectively.  Estimated metallicities (gray line) are determined from the LGRB host mass and redshift data using the mass-metallicity-redshift relation shown in \aref{Zahid_polyfit}. A green box with a width of 0.25 dex is overplotted; this corresponds to a difference factor of 1.8 in linear scale.  This resolves much of the difference between the LGRB formation metallicity cutoff of about a third solar in \cite{diff_rate_letter} with the cutoff value of approximately solar claimed in \cite{Perley_shoals_masses} in favor of the former.} 

\end{minipage}
\begin{minipage}[t]{0.03\textwidth}
~
\end{minipage}
\begin{minipage}[t]{0.45\textwidth}

\includegraphics[width=1\textwidth,height=1\textwidth]{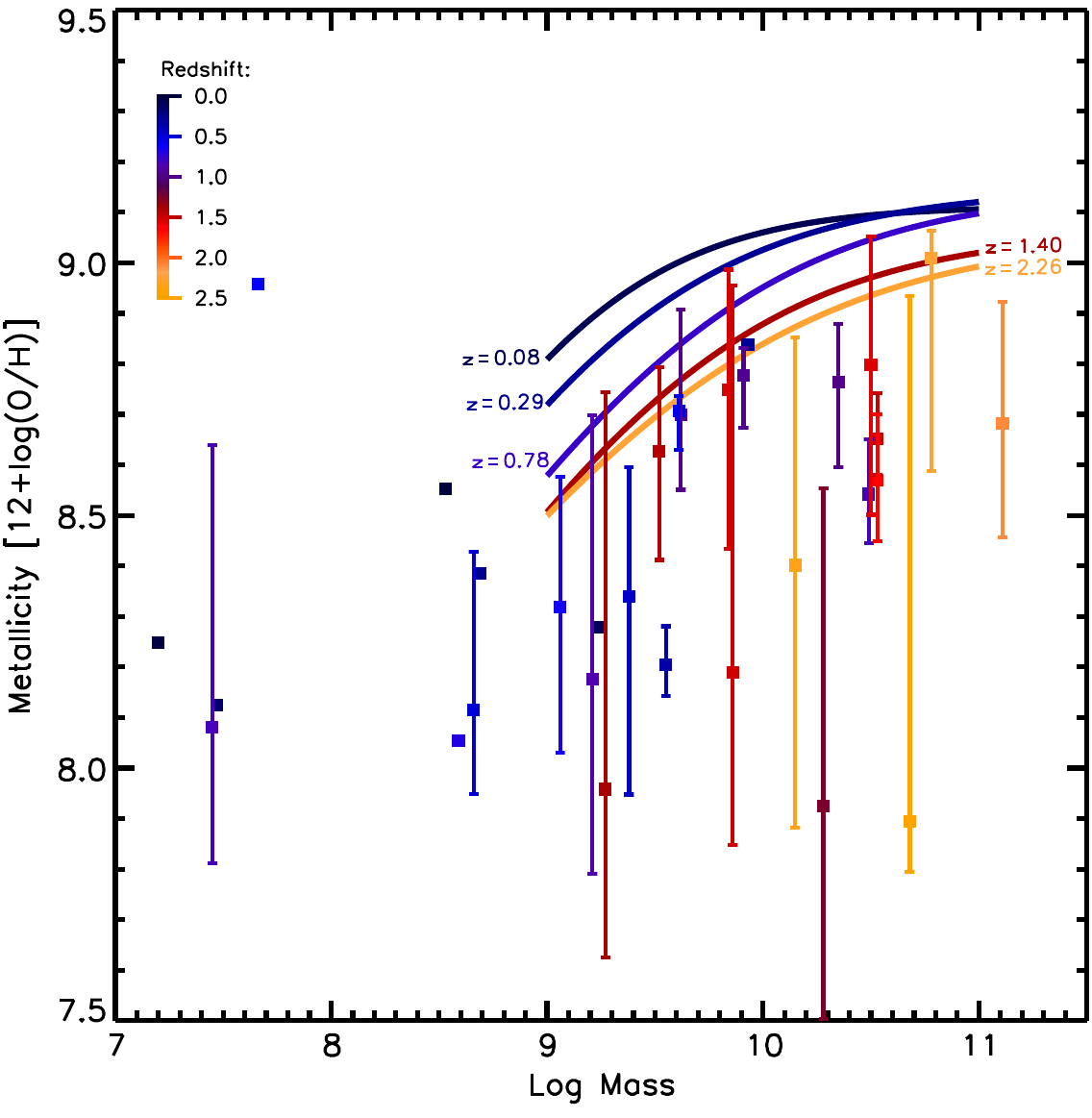}
\caption{\label{metallicity_vs_mass} Metallicity vs.\ mass for objects with both measured and estimated metallicity values (i.e.\ those given in \aref{metal_comp_table}) with our fitting of the mass metallicity relation curves of  \cite{Zahid2013} shown for comparison.  The redshifts of the different objects and comparison curves is indicated by color. Metallicity errors are shown for objects with errors in \aref{xsd_data_table}. Objects without metallicity errors do not have line fluxes with error measurement suitable to estimate errors.  Mass errors are not given in the samples from which the mass values are compiled nor are errors given in the raw photometry from which they were determined.}
\end{minipage}
\end{center}
\end{figure*}

Comparing \aref{shoals_metal_est_step_plot_5} to \aref{measured}, we see that the MZR-based metallicity distributions have higher metallicities for a given redshift than the spectroscopically measured metallicity distributions. We show this explicitly in \aref{measured_vs_estimated}, where we plot the measured and estimated metallicity distributions (simplified to just a single 0~$<$~{\it z}~$<$~2.5 redshift bin) next to each other. The offset between the two distributions is, on average, approximately a quarter dex, and is roughly constant across the distributions interquartile range. Therefore, assuming that the mass values are not overestimated (which may be the case), the LGRB host galaxy population is systematically lower in metallicity than typical galaxies of comparable mass and redshift. This result is unlikely to be a result of the systematic offsets known to exist between metallicity diagnostics, since \cite{Zahid2013} used the same \citetalias{KobulnickyKewley} R$_{23}$ diagnostic as we use here. The offset in the predicted and measured metallicities thus suggests that the LGRBs are biased towards the lowest metallicities within any galaxy population, as has previously been found \citep[e.g.,][]{Savaglio,stats_paper,xshooter_survey,diff_rate_letter,Palmerio2019}, and can not be correctly modeled using the general mass metallicity relation. Evidence of this can be seen in \aref{metallicity_vs_mass}, where we show a scatterplot of the subset of objects for which we have both measured metallicities and masses with our fitting of the mass metallicity relation curves of \cite{Zahid2013} overplotted for comparison. The mass-metallicity curves are generally at high metallicity (above solar at $z<0.5$), which is an effect of the \citetalias{KobulnickyKewley} R$_{23}$ diagnostic used, which is known to produce systematically higher metallicities than empirically-derived metallicity diagnostics \citep[e.g.,][]{KewleyEllison,AndrewsMartini}. However, it is the relative difference in metallicities rather than the absolute metallicity value that is of importance here.

In \aref{metal_comp_table} we explicitly compare the 29 objects for which we have both measured and estimated metallicity values. We find a mean difference between the estimated and measured metallicities of 0.23 dex with a median difference of 0.30 dex but with a standard deviation on the difference of 0.46 dex. In raw (i.e.\ non-log) metallicities we find a median ratio of 0.50 between the measured and estimated metallicities. Hence we conclude that while the estimated metallicities are typically higher than the measured values, the scatter between measured and mass-based metallicities is too large to devise a correction that will bring the two into agreement.

The possibility exists that the \cite{Perley_shoals_masses} SHOALS mass values are exaggerated as stellar mass values derived from a single NIR observation can significantly overestimate the mass of galaxies \citep{Heintz2018, Corre2018, Arabsalmani2018}.  In particular, \cite{Heintz2018} cites a difference of 0.2 to 0.3 dex between values in SHOALS and stellar masses obtained by fitting the same host galaxy SED, but using multiple optical and NIR photometric bands. If the stellar masses are systematically inflated, then the estimated metallicities would also be high. From \aref{Zahid_polyfit}, a quarter dex difference in mass would suggest differences of about 0.1 dex in metallicity which is not enough to explain the quarter dex difference between the measured and estimated metallicities we find here (see \aref{measured_vs_estimated}).

\section{Speculative Discussion of Possible Causes}\label{Discussion}
The absence of evolution in the LGRB metallicity distribution is quite puzzling and does not seem to conform to our general expectations of galaxy enrichment. While we cannot quantify the impact of selection effects on the sample of host galaxies with good quality spectra, the fact that we also find an absence of evolution with redshift in the metallicities that we estimate from the host galaxy mass, of which very complete samples exist, suggests that this result is not the product of selection biases.  Nor would the product of selection biases have reason to particularly favor a metallicity distribution that remains constant with redshift. We therefore proceed on the assumption that the observed constant metallicity distribution is real, and discuss the possible origins and implications.

The simplest explanation of the known LGRB preference for low metallicity environments \citep{LeFloch2006,Stanek2007, Fruchter,Savaglio,stats_paper, diff_rate_letter, Perley_shoals_masses, Levesque051022, Levesque2} is that this effect is caused by a difference in the LGRB formation rate per unit of underlying star-formation at different metallicities (c.f.\ \citealt{diff_rate_letter}, see footnote\footnote{All else being equal, the LGRB rate should uniformly follow the star-formation rate as LGRBs are the product of massive short-lived progenitor stars, which are assumably created as a uniform fraction of the overall star-formation.  \citealt{diff_rate_letter} contrasts this assumption with the observed metallicity distributions of the star-formation rate and of LGRB hosts, finding that the observed LGRB host metallicity distribution requires a roughly $\sim$30 fold suppression in LGRB formation rate at metallicities above 12+log(O/H) $\gtrapprox$ 8.3 (\citetalias{KobulnickyKewley} scale) as compared with the underlying star-formation rate to LGRB formation rate efficiency at lower metallicities.}), as predicted by the collapsar model \citep{Woosley1993}. However, this scenario alone does not explain the notable fraction of detected super-solar LGRB host galaxies, nor does it account for the constant fraction of high-metallicity host galaxies with redshift. In fact, given the decrease in the cosmic chemical enrichment with redshift, a preference for metal-poor progenitors should manifest itself as a decrease in the fraction of metal-rich LGRB host galaxies with redshift.

One possible explanation that may account for this apparent discrepancy is that the high metallicity LGRB hosts are not representative of the general galaxy population that make up the samples used to study the MZR at high redshift, which generally rely on stacked spectra \cite[e.g.,][]{Erb2006,Yabe2012}. LGRB host galaxies may, instead, be in a short-lived transition phase, having recently been low metallicity galaxies that have undergone a sudden burst of enrichment.  This could also explain the existence of high metallicity LGRBs, since the gas that formed the progenitor may have been segregated from this enrichment process, thus allowing the LGRB progenitor to form in a low metallicity environment, but the LGRB itself to occur in a metal-enriched environment.

There is evidence that a higher fraction of Type Ibc SNe are found in merging galaxies than in the case of Type II SNe\citep{Hakobyan2014}, and there is also an indication of an elevated fraction of mergers or interacting systems among LGRB host galaxies \citep{Chen_HW2012,Savaglio2012}. Furthermore, some simulations suggest that the increased star formation triggered by galaxy interactions can result in a rapid increase in gas-phase metallicity of up to 1.5~dex/Gyr \citep{Torrey2012}, although this phase of rapid enrichment is short-lived, and is soon followed by dilution from the efficient inflow of metal-poor gas that is channelled along torques formed during the galaxy interaction \citep{Perez2011, Torrey2012}. The constant fraction of metal rich LGRB host galaxies with redshift may thus be capturing a predicted, short-lived phase of galaxy enrichment as galaxies interact, thus not reflecting the metallicity of the LGRB progenitor itself, which formed from less enriched material. \cite{Perez2011} also found that the enhancement in merger-induced star formation and oxygen abundance was dependent on the metallicity and gas mass of the two interacting galaxies, which may help explain why the fraction of metal-rich LGRB host galaxies that we find remains constant up to $z=2.5$. On the other hand, there are observations that indicate that galaxy pairs have lower metallicity to field galaxies, which has been argued to be a result of metallicity dilution by strong inflows of metal-poor gas \cite{Rupke2010,Ellison2013,Gronnow2015}. It is possible that the enrichment or dilution of galaxies during or shortly after a merger will depend on the mass of the galaxies involved, and on the phase of the merger history, as well as the redshift of the system \citep{Michel-Dansac2008,Sparre2022}. It it therefore not possible to reach any strong conclusions on what may be the origin of the constant, high metallicity fraction that we see in the LGRB host galaxy population. Greater samples of LGRB host galaxies with well-characterized morphologies are needed to explore this hypothesis further, as well as further theoretical work on how mergers may affect the rate of LGRBs, and subsequent host galaxy metallicity.

\section{Conclusions}

In this paper we have taken a sample of 55 LGRB host galaxy spectra to study the evolution in the metallicity distribution of LGRB host galaxies out to $z<2.5$. We find a surprising lack of evolution in the metallicity distribution of our LGRB host galaxy sample with redshift (section~\ref{mzd}), although we can't fully quantify the effect that selection effects may be having on this result. However, we note that from comparisons with other LGRB host galaxy samples with more clearly defined selection functions, we find little evidence for strong selection effects in our sample in terms of the redshift distribution (section~\ref{Host_redshifts}), or in our sample's host galaxy metallicity, stellar mass, or SFR distribution (section \ref{Selection_Effects}).

Nevertheless, given the likely complex selection effects that are present in our sample, we focus our results on the more robust and surprising finding that the fraction of metal-rich (12+log(O/H)$>$8.4) LGRB host galaxies remains constant with redshift (relative to the full sample of LGRBs with known redshift; see \aref{Z_frac}). While we cannot exclude the possibility of Malmquist bias contributing to our results, we would not expect Malmquist bias to produce an unevolving fraction of metal-rich LGRB host galaxies.

In \aref{Discussion} we hypothesize that LGRBs found in metal-rich host galaxies may be selecting galaxies in a short-lived phase of a gas-rich merger, during which time an increase in star formation results in the rapid enrichment of the interstellar medium. Since wet-mergers were more prevalent in the younger Universe \citep[e.g.,][]{Lin,Chou}, this explanation could account for the apparent large fraction of metal-rich GRB host galaxies at high redshift. Nevertheless, further theoretical investigation is required to demonstrate that such a model would indeed produce a constant fraction of metal-rich LGRB host galaxies, and more observational data is needed to firmly quantify the fraction of galaxy mergers in the population high-z, metal-rich LGRB host galaxies (relative to the fraction of mergers in the general, star-forming galaxy population). With the successful launch of the James Webb Space Telescope (JWST), and the preliminary indications of its superb sensitivity, it will soon be possible to obtain accurate merger fractions of star forming galaxies at $z>2$, including LGRB host galaxies, making it possible to test this hypothesis further.

An alternative, but very sensitive method of tracing the chemical composition of distant galaxies is through absorption line metallicities. LGRB host galaxies themselves are uniquely suited to absorption line metallicity measurements, as the GRB afterglow itself provides a bright background source clean of intrinsic spectral features. Furthermore, LGRB sight lines generally pierce through the central star forming regions of their host galaxy \citep{Fruchter, Blanchard}, and the observed absorption lines thus trace the material within the galaxy disk. Afterglow absorption line metallicities are particularly sensitive at low metallicities, where attenuation by dust is less significant \citep{DeCia2013,DeCia2016,Wiseman,Bolmer2019}, and they therefore offer a complementary tool to emission line diagnostics, which are generally biased in favour of more luminous and metal rich galaxies.
Nevertheless, whereas emission line metallicities are luminosity weighted, and thus trace the ionized gas within star forming regions, LGRB absorption line metallicities trace the neutral gas phase of the interstellar medium. It therefore remains to be verified whether absorption and emission line metallicities are sufficiently comparable to allow cross-calibration, and thus whether they can be uniformly combined to probe the cosmic chemical evolution in a consistent manner out to lower metallicities and higher redshifts.

The unprecedented sensitivity of JWST at near-infrared wavelengths will soon make it possible to measure the emission line oxygen abundance of typical star forming galaxies out to $z>4$. Also, the astonishing detection of the temperature-sensitive [\ion{O}{3}]$\lambda$4363 emission line in the NIRSpec spectrum of an $z=8.5$ galaxy released as part of the JWST first images offers great hope in the prospect of developing strong emission line metallicity diagnostics that are calibrated directly to the conditions of high-$z$ galaxies. To this end, as part of a larger collaboration, we are undertaking an effort using JWST to measure emission line metallicities for a sample of LGRB host galaxies with precise, absorption line metallicities already in hand, thus offering an opportunity to cross-calibrate these two metallicity tracers.  With JWST performing better than expected, it should soon be possible to extend the galaxy mass-metallicity relation out to further redshifts and lower-mass galaxies, thus offering a more detailed understanding of how metallicity shapes the formation process of the LGRBs seen in high metallicity host galaxies and thus how LGRBs form in general.

\begin{acknowledgments}
We thank Thomas Kr{\"u}hler for many useful discussions and assorted assistance.  His omission as a coauthor is due solely to his personal preference since leaving academia.  His presence in astronomy is missed and we wish him well in his new career.

We thank the BAT6 team for a detailed explanation of their {\it z} $\gtrsim$ 2.5 selection effects.  We also acknowledge and thank Dan Perley and our anonymous referees for a number of helpful suggestions, in particular our final referee, Sandra Savaglio (who waved anonymity after acceptance), for detailed comments and feedback that helped improve the quality of the paper.


John Graham acknowledges support through the National Science Foundation of China (NSFC) under grant 11750110418.

John Graham also offers profound thanks to the University of Hawaii at Manoa Department of Physics, in particular John Learned and Peter Gorham, for their hospitality and assistance after a period of prolonged illness.

\end{acknowledgments}

\appendix
\vspace{-0.5 cm}\twocolumngrid
\renewcommand{\floatpagefraction}{1}
\setcounter{table}{0}
\renewcommand{\thetable}{A\arabic{table}}
\begin{sidewaystable*}[p!]
\begin{center}
\begin{minipage}[p]{1.07\textwidth}
\caption{\label{LGRB_num_stat_KS_table} K-S probabilities for \aref{redshift_distribution}}
\end{minipage}
\footnotesize
\renewcommand{\baselinestretch}{0.80}\selectfont
\hspace{-0.8 cm}\begin{tabular}{@{\hskip -0.2 cm}m{4.4cm}|L{1.2cm}L{1.2cm}L{1.2cm}L{1.2cm}L{1.2cm}L{1.2cm}L{1.2cm}L{1.2cm}L{1.2cm}L{1.2cm}L{1.2cm}L{1.2cm}L{1.2cm}}
\hline
\hline
 & All LGRBs Redshifts & All LGRBs Redshifts z~$<$~2.5 & GF13 Metalicites & GF13 Metalicites z~$<$~0.5 & GF13 Metalicites z~$>$~0.5 & Svensson Masses & Kruehler Redshifts & Kruehler Redshifts z~$<$~2.5 & Kruehler Metalicites & Kruehler Metalicites z~$<$~2 & Kruehler Metalicites z~$>$~2 & Kruehler Metalicites Z $>$ 8.7\\
\hline
All LGRBs Redshifts z~$<$~2.5 &            1\\
GF13 Metalicites &     0.1748 &     0.1748\\
GF13 Metalicites z~$<$~0.5 &     0.6184 &     0.6184 &            1\\
GF13 Metalicites z~$>$~0.5 &     0.9806 &     0.9806 &            1 &       -NaN\\
Svensson Masses &     0.7626 &     0.7626 &     0.3700 &     0.9748 &     0.9119\\
Kruehler Redshifts &     0.5693 &     0.8064 &     0.0371 &     0.3129 &     0.9518 &     0.9432\\
Kruehler Redshifts z~$<$~2.5 &     0.8064 &     0.8064 &     0.0371 &     0.3129 &     0.9518 &     0.9432 &            1\\
Kruehler Metalicites &     0.7898 &     0.7898 &     0.1341 &     0.3713 &     0.6024 &     0.6950 &     0.7510 &     0.7510\\
Kruehler Metalicites z~$<$~2 &     0.7404 &     0.7404 &     0.1341 &     0.3713 &     0.6024 &     0.6950 &     0.9989 &     0.9989 &            1\\
Kruehler Metalicites z~$>$~2 &     0.8379 &     0.8379 &       -NaN &       -NaN &       -NaN &       -NaN &     0.9813 &     0.9813 &            1 &       -NaN\\
Kruehler Metalicites Z $>$ 8.7 &     0.9565 &     0.9565 &     0.3000 &       -NaN &       -NaN &     0.9096 &     0.9445 &     0.9445 &     0.9974 &     0.9975 &       -NaN\\
Kruehler Metalicites Z $>$ 8.4 &     0.8965 &     0.8965 &     0.0862 &       -NaN &       -NaN &     0.6746 &     0.9970 &     0.9970 &     0.9991 &     1.0000 &            1 &     1.0000\\
Kruehler Metallicities Z $>$ 8.7 \& z~$<$~2 &     0.6429 &     0.6429 &     0.3000 &       -NaN &       -NaN &     0.9096 &     0.9230 &     0.9230 &     0.9659 &     0.9659 &       -NaN &            1\\
Kruehler Metallicities Z $>$ 8.4 \& z~$<$~2 &     0.3951 &     0.3951 &     0.0862 &       -NaN &       -NaN &     0.6746 &     0.9496 &     0.9496 &     1.0000 &     1.0000 &       -NaN &     1.0000\\
Kruehler Metallicities Z $>$ 8.4 \& z~$>$~2 &     0.9226 &     0.9226 &       -NaN &       -NaN &       -NaN &       -NaN &     0.9956 &     0.9956 &            1 &       -NaN &            1 &       -NaN\\
TOUGHS Redshifts &     0.0237 &     0.1464 &     0.2031 &     0.9990 &     0.9969 &     0.8574 &     0.0403 &     0.4891 &     0.1895 &     1.0000 &     0.9842 &     0.9169\\
TOUGHS Redshifts z~$<$~2.5 &     0.1464 &     0.1464 &     0.2031 &     0.9990 &     0.9969 &     0.8574 &     0.4891 &     0.4891 &     0.1895 &     1.0000 &     0.9842 &     0.9169\\
BAT6 sample &     0.6498 &     0.3261 &     0.0377 &       -NaN &     0.9969 &     0.8850 &     0.8316 &     0.8869 &     0.4225 &     0.9732 &     0.1350 &     0.3417\\
BAT6 sample z~$<$~2.5 &     0.3261 &     0.3261 &     0.0377 &       -NaN &     0.9969 &     0.8850 &     0.8869 &     0.8869 &     0.4225 &     0.9732 &     0.1350 &     0.3417\\
SHOALS Masses &     0.0866 &     0.0847 &     0.0116 &       -NaN &     0.9886 &     0.6808 &     0.1805 &     0.6300 &     0.1443 &     0.8389 &     0.6554 &     0.2994\\
SHOALS Masses z~$<$~2.5 &     0.0847 &     0.0847 &     0.0116 &       -NaN &     0.9886 &     0.6808 &     0.6300 &     0.6300 &     0.1443 &     0.8389 &     0.6554 &     0.2994\\
SHOALS Redshifts &     0.0158 &     0.0355 &     0.0102 &     0.4746 &     0.9543 &     0.6901 &     0.0904 &     0.5442 &     0.0974 &     0.7849 &     0.8124 &     0.3332\\
SHOALS Redshifts z~$<$~2.5 &     0.0355 &     0.0355 &     0.0102 &     0.4746 &     0.9543 &     0.6901 &     0.5442 &     0.5442 &     0.0974 &     0.7849 &     0.8124 &     0.3332\\
GS16 Prediction &     0.0000 &     0.0007 &     0.0592 &     0.4512 &     0.9944 &     0.5836 &     0.0137 &     0.5849 &     0.1176 &     0.2831 &     0.9923 &     0.5382\\
GS16 Prediction z~$<$~2.5 &     0.0007 &     0.0007 &     0.0592 &     0.4512 &     0.9944 &     0.5836 &     0.5849 &     0.5849 &     0.1176 &     0.2831 &     0.9923 &     0.5382\\

\\ \hline
\hline
 & Kruehler Metalicites Z $>$ 8.4 & Kruehler Metallicities Z $>$ 8.7 \& z~$<$~2 & Kruehler Metallicities Z $>$ 8.4 \& z~$<$~2 & Kruehler Metallicities Z $>$ 8.4 \& z~$>$~2 & TOUGHS Redshifts & TOUGHS Redshifts z~$<$~2.5 & BAT6 sample & BAT6 sample z~$<$~2.5 & SHOALS Masses & SHOALS Masses z~$<$~2.5 & SHOALS Redshifts & SHOALS Redshifts z~$<$~2.5 \\
\hline
Kruehler Metallicities Z $>$ 8.7 \& z~$<$~2 &     0.9987\\
Kruehler Metallicities Z $>$ 8.4 \& z~$<$~2 &            1 &     0.9987\\
Kruehler Metallicities Z $>$ 8.4 \& z~$>$~2 &            1 &       -NaN &       -NaN\\
TOUGHS Redshifts &     0.6018 &     0.9987 &     0.9982 &     0.8546\\
TOUGHS Redshifts z~$<$~2.5 &     0.6018 &     0.9987 &     0.9982 &     0.8546 &            1\\
BAT6 sample &     0.8309 &     0.8053 &     0.8672 &     0.2396 &     0.0479 &     0.6518\\
BAT6 sample z~$<$~2.5 &     0.8309 &     0.8053 &     0.8672 &     0.2396 &     0.6518 &     0.6518 &            1\\
SHOALS Masses &     0.4681 &     0.9963 &     0.9497 &     0.7533 &     0.8041 &     0.9859 &     0.3573 &     0.9841\\
SHOALS Masses z~$<$~2.5 &     0.4681 &     0.9963 &     0.9497 &     0.7533 &     0.9859 &     0.9859 &     0.9841 &     0.9841 &            1\\
SHOALS Redshifts &     0.4210 &     0.9866 &     0.9374 &     0.9362 &     0.9739 &     0.9991 &     0.1837 &     0.9593 &     0.9991 &     1.0000\\
SHOALS Redshifts z~$<$~2.5 &     0.4210 &     0.9866 &     0.9374 &     0.9362 &     0.9991 &     0.9991 &     0.9593 &     0.9593 &     1.0000 &     1.0000 &            1\\
GS16 Prediction &     0.4486 &     0.7800 &     0.5734 &     0.9712 &     0.1776 &     0.1034 &     0.0739 &     0.8843 &     0.6397 &     0.3258 &     0.6038 &     0.1711\\
GS16 Prediction z~$<$~2.5 &     0.4486 &     0.7800 &     0.5734 &     0.9712 &     0.1034 &     0.1034 &     0.8843 &     0.8843 &     0.3258 &     0.3258 &     0.1711 &     0.1711 \\

\\ \hline
\end{tabular}
\end{center}
\vspace{-0.2 cm}
\hspace{-0.4 cm}
\begin{minipage}[H]{1.03\textwidth}
{\vspace{0.0 cm}\normalsize Computed Kolmogorov--Smirnov (K-S) probabilities comparing the lines in the \aref{redshift_distribution} redshift distributions (and some additional subsets thereof). This was affected by, for each paring, removing the objects of each line outside the redshift range of the other (with a grace of 0.05 $z$) and then running a normal K-S test on the remaining values. NaN values indicate K-S test failure due to having fewer than 4 objects in a redshift matched comparison sample.  Values of ``1" are exact and indicate a sample which, due to redshift range matching cuts, is being evaluated with itself, values of ``1.0000" are the result of rounding.  Note: The K-S values in this table are not dependent on the line scaling factors used in \aref{redshift_distribution}.}
\end{minipage}
\end{sidewaystable*}

\begin{sidewaystable*}[p!]
\begin{center}
\begin{minipage}[H]{0.55\textwidth}
\vspace{1 cm}\caption{\label{redshift_scaling} Number of objects per redshift bin in \aref{redshift_distribution} and scalings used for each of the comparison samples to normalize each distribution to approximately 60 at $z=2.5$.}
\end{minipage}
\footnotesize
\begin{tabular}{m{4.1cm}|L{1.4cm}L{1.4cm}L{1.4cm}L{1.4cm}L{1.4cm}L{1.4cm}L{1.4cm}L{1.4cm}L{1.4cm}L{1.4cm}L{0.9cm}}
\hline
\hline
 & 0.0~$<$~z~$<$~0.5 & 0.5~$<$~z~$<$~1.0 & 1.0~$<$~z~$<$~1.5 & 1.2~$<$~z~$<$~1.7 & 1.5~$<$~z~$<$~2.0 & 2.0~$<$~z~$<$~2.5 & 2.1~$<$~z~$<$~2.5 & 2.5~$<$~z~$<$~3.0 & 3.0~$<$~z~$<$~3.5 & 3.5~$<$~z~$<$~4.0 & Scaling factor\\ \hline
All known LGRBs w/ redshifts & 64 & 91 & 82 & 78 & 65 & 76 & 55 & 40 & 33 & 21 & 0.2 \\
BAT6 sample w/ redshifts & 3 & 11 & 9 & 11 & 10 & 11 & 9 & 5 & 2 & 2 & 1.3 \\
TOUGH sample w/ redshifts & 4 & 8 & 7 & 9 & 9 & 11 & 9 & 6 & 6 & 6 & 1.5 \\
SHOALS sample w/ redshifts & 4 & 18 & 13 & 17 & 17 & 22 & 15 & 10 & 11 & 6 & 0.8 \\
SHOALS sample w/ M$_\star$ & 3 & 15 & 9 & 13 & 14 & 16 & 9 & 7 & 8 & 2 & 1.1 \\
Svensson+10 w/ M$_\star$ & 10 & 17 & 7 & 0 & 0 & 0 & 0 & 0 & 0 & 0 & 0.1 \\
Kr{\"u}hler+15 LGRBs w/ redshifts & 10 & 23 & 16 & 19 & 16 & 21 & 17 & 6 & 4 & 1 & 0.7 \\
Kr{\"u}hler+15 LGRBs w/ metallicities & 6 & 15 & 8 & 14 & 6 & 9 & 8 & 0 & 0 & 0 & 1 \\
Kr{\"u}hler+15 LGRBs w/ $Z>Z_\odot$ & 2 & 5 & 3 & 5 & 2 & 2 & 2 & 0 & 0 & 0 & 3 \\
Kr{\"u}hler+15 LGRBs w/ $Z>0.5Z_\odot$ & 3 & 11 & 5 & 9 & 4 & 8 & 7 & 0 & 0 & 0 & 1.5 \\
GF13 sample & 10 & 4 & 0 & 0 & 0 & 0 & 0 & 0 & 0 & 0 & 0.7 \\
GS16 Prediction & 222 & 437 & 579 & 517 & 487 & 398 & 312 & 358 & 289 & 254 & - \\

\hline
\end{tabular}
\end{center}
\hspace{0.1 cm}\begin{minipage}[H]{0.99\textwidth}
{\vspace{-0.3 cm}\normalsize Note the irregular and overlapping bins from 1.2~$<$~{\it z}~$<$~1.7 and 2.1~$<$~{\it z}~$<$~2.5  to avoid the redshift gaps (from 1.0~$\lesssim$~{\it z}~$\lesssim$~1.2 and 1.7~$\lesssim$~{\it z}~$\lesssim$~2.1) in the \cite{xshooter_survey} metallicity sample.  Also note that the number of simulated objects in the GS16 \citep{form_rate_letter} prediction is arbitrary, however the ratio of the number of objects between the different redshift bins is meaningful.}
\end{minipage}

\end{sidewaystable*}

\section{Redshift comparison samples}\label{AppendixA}

Details on each of the LGRB host galaxy samples plotted in \aref{redshift_distribution} are given below. In \aref{LGRB_num_stat_KS_table} we also provide the K-S test p-values for each pair of curves in \aref{redshift_distribution}, which gives the probability that the null hypothesis (that samples are related) is true. It is standard to use a p-value $<0.05$ in order to reject the null hypothesis, which would indicate that the two samples being compared are significantly different, and are thus likely not taken from the same parent population. In \aref{redshift_scaling} we summarize the number of sources within each comparison sample shown in \aref{redshift_distribution} at different redshift bins, and also give the scaling factors used to normalize each sample redshift distribution.

\begin{itemize}
\item We begin by plotting the redshift distribution of the LGRB host galaxy population with host metallicity measurements taken from \cite{xshooter_survey}, corresponding to 44\% of the parent sample. In addition we show the full sample from \cite{stats_paper}, which is scaled by a factor of 0.7 to match the distribution of the \cite{xshooter_survey} metallicity sub-sample.

\item To address concerns about a metallicity dependent redshift bias we also plot the subset of \cite{xshooter_survey} hosts with metallicities above solar (16\% of the parent sample), and scale the distribution by a factor of three.  We also plot the subset of \cite{xshooter_survey} hosts with metallicities above log(O/H)+12 $> 8.4 $ (\citetalias{KobulnickyKewley} scale) (31\% of the parent sample) so as to sample only objects clearly above the log(O/H)+12 $\approx 8.3 $ metallicity cutoff noted in \cite{diff_rate_letter}. This latter distribution is scaled by a factor of 1.5 in \aref{redshift_distribution}.

\item To address concerns about a bias in which objects have the lines needed for metallicity measurement we plot the entire \cite{xshooter_survey} host population including the objects without a metallicity.  Since metallicity measurement requires a specific set of lines whereas redshift measurement requires only a single identified line (in practice any two lines is used to identify both the lines and the redshift) this provides an immediate crosscheck on observational biases in line measurement. Including \cite{xshooter_survey} objects with redshift but not metallicity makes an approximate 50\% increase in the sample size (at $z$~$<$~2.5 before the slope noticeably changes), and this sample is thus scaled down by a factor of 0.67.  As the slope of this, and some other distributions, noticeably changes at $z$~$\approx$~2.5, and since $z$~$\approx$~2.5 is the limit of the redshift range for which we have measured metallicities, we adopt this redshift, along with the matching ordinate of this distribution (a value of $\sim$60), as the reference point for scaling the remaining distributions.

\item We also include the full sample of known LGRB host galaxy redshifts taken from the GRB table of Jochen Greiner\footnote{\url{http://www.mpe.mpg.de/~jcg/grbgen.html}} (excluding for this and all other samples those objects with no or only photometric redshifts, { and of course removing the short-bursts to ensure an LGRB-only sample}), as of UTC noon March 15\textsuperscript{th} 2019 (25\% of the sample).  As this sample is undoubtedly subject to a variety of complex selection effects, its use provides a reasonable assessment of the degree that such unmitigated effects have on our analysis.

\item To address biases in host redshift measurement as a function of redshift itself { (i.e.\ redshift dependent selection effects on which objects receive redshift measurement)}, we plot the \cite{Perley_shoals_masses} SHOALS (88\% complete in redshift) and \cite{BAT6} BAT6 samples (88\% complete in redshift), which claim to be unbiased in their selections.  We also plot the Optically Unbiased Gamma-Ray Burst Host Survey (TOUGH) sample updated using all currently known spectroscopically determined redshift values for objects from that sample. TOUGH is an attempt to provide a host sample which is not biased by host absorption of the optical afterglow of the LGRB. Of the 69 {\it Swift} GRB hosts we find 58 of them have exact redshift values.  We note that one of the missing redshifts is due to a bright foreground star and can thus be excluded from the sample without introducing any biases, giving the TOUGH redshift sample an 85\% effective completeness.

\item To also sample bias in host galaxy mass measurement we plot the subset of \cite{Perley_shoals_masses} SHOALS galaxies with mass values (excluding limits) (69\% of the sample) as well as the sample of \cite{Svensson}.

\end{itemize}

We also consider a theoretical curve for the LGRB event rate. \cite{form_rate_letter} estimated the LGRB progenitor rate using the \cite{diff_rate_letter} results in concert with SNe statistics via an approach patterned loosely off the Drake equation.  Beginning with the cosmic star-formation history, \cite{form_rate_letter} took the expected number of broad-line Type Ic events that are in low metallicity host environments as potential LGRB progenitor candidates.  They then adjusted this number by adding the contribution of broad-line Type Ic SNe in high metallicity host environments at a much reduced weighting of $\sim$$\frac{1}{30}$ (the \citealt{diff_rate_letter} estimate on the relative suppression of LGRB formation in high metallicity environments).  A comparison of this estimate of potential LGRB progenitor candidates to the observed LGRB rate corrected for instrumental selection effects (estimated by \citealt{Lien} and \citealt{Graff}), provided a combined estimate of the fraction of broad-line Type Ic SNe residing in metal poor environments that produce an LGRB and the fraction of such LGRBs that are beamed in our direction.
Thus \cite{form_rate_letter} estimated that, in a low metallically environment, an aligned LGRB occurs out of approximately every 4000 $\pm$ 2000 broad-lined Type Ic Supernovae.  Therefore if one assumes a semi-nominal beaming factor of 100 then 1 out of $\sim$40 low metallicity Ic-bl SNe give rise to an LGRB.

Using the process described above, we derive an expected LGRB event rate as follows: Beginning with the cosmic star-formation rate history of \cite{hb} and the \cite{form_rate_letter} estimates of the metallicity distribution of the universe as a function of redshift we obtain an estimate of the amount of star-formation above and below the \cite{diff_rate_letter} metallicity cutoff.  The expected number of LGRBs formed is thus estimated after applying the \cite{diff_rate_letter} estimate of the effect of metallicity on the LGRB formulation rate per unit of available star-formation.  The result is an expected LGRB event rate as a function of redshift which can then be scaled in the same manner as the observed LGRB populations described previously.

\newcommand{\CACI}{\citetalias{Curti2017}} 
\newcommand{\CCI}{\citealt{Curti2017}} 
\newcommand{\CCII}{\citealt{Curti2020}} 

\section{\CACI~(\CCI~optimized via \CCII) metallicity diagnostic}
\label{CCM+17}\label{AppendixB}
\cite{Curti2017} simultaneously (re)calibrated the six strong emission line ratio relations given below against temperature, or $T_e$-based metallicities, using a sample 118~478 galaxies from SDSS DR7 with a median redshift $z=0.072$. 


\setcounter{figure}{0}
\renewcommand{\thefigure}{B\arabic{figure}}
\begin{figure*}[t]
\begin{center}
\includegraphics[width=0.49\textwidth]{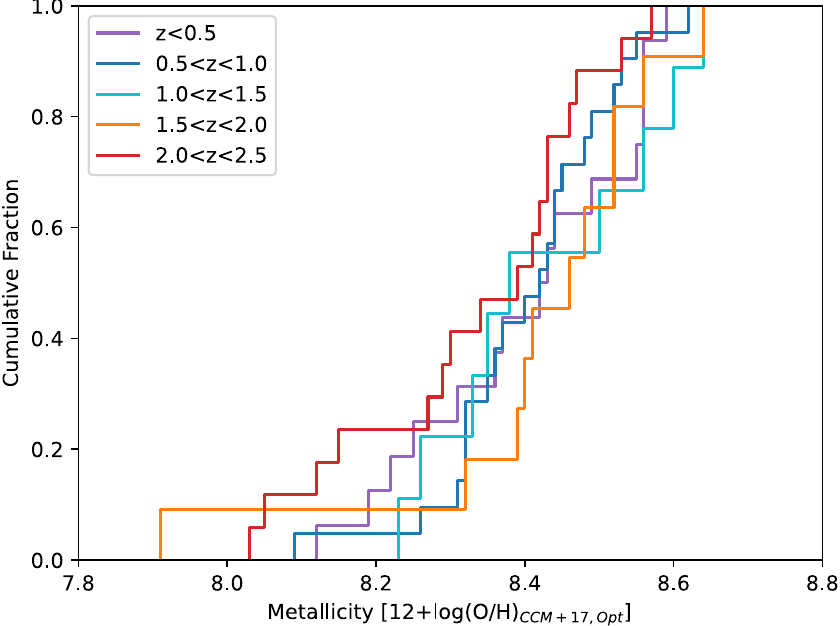}
\includegraphics[width=0.49\textwidth]{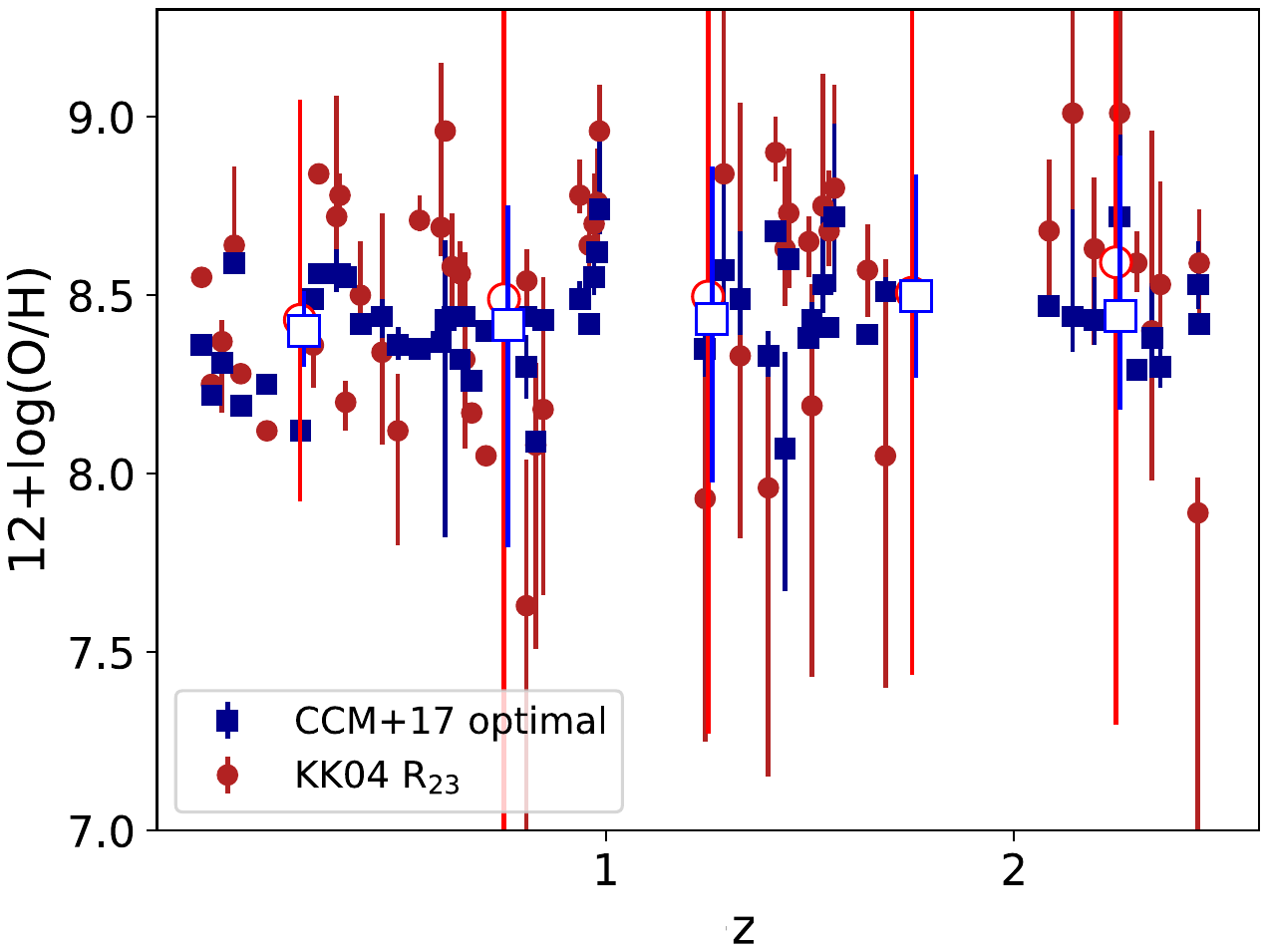}

\caption{\label{BF_metallicity_CDFs} 
Cumulative distribution similar figure to \aref{measured} (left pane) and a similar figure to \aref{metallicity_vs_redshift} (right pane), but with the measured LGRB host metallicities now derived using the metallicity diagnostic calibrations from \cite{Curti2017} and method described in \cite{Curti2020}. In the right pane the KK04 \citep{KobulnickyKewley} diagnostic results have been overplotted. The average R$_{23}$ and \citetalias{Curti2017} optimal metallicities in bins of $\Delta z =0.5$ are also plotted as larger, open, circle and square symbols, respectively, which illustrate the lack of evolution in metallicity with redshift for both metallicity diagnostics. We have offset the average metallicity values by 0.01dex for clarity. Many of the data points at $z<1$ correspond to LGRB host galaxies taken from \citet{stats_paper}, which have no reported line flux measurements, and we therefore report no uncertainty on the derived metallicities. Note that since the metallicity method of \cite{Curti2020} includes multiple metallicity diagnostics, with each using a different set of emission line ratios, this method provides metallicities for a larger sample of LGRB host galaxies (79) than when applying the R$_{23}$ diagnostic alone, which requires the detection of \ion{O}{2}, \ion{O}{3} and H$\beta$. The \cite{Curti2020} metallicities for those GRBs not listed in \aref{xsd_data_table} are given in \aref{xsd_CCM17_data_table}.}
\end{center}
\end{figure*}

\begin{itemize}\label{emission line ratio relations}
\item[] R$_2$=[\ion{O}{2}]$\lambda 3727$/H$\beta$
\item[] R$_3$=[\ion{O}{3}]$\lambda 5007$/H$\beta$
\item[] R$_{23}$=([\ion{O}{2}]$\lambda 3727$+[\ion{O}{3}]$\lambda\lambda 4959,5007$)/H$\beta$
\item[] O$_{32}$=[\ion{O}{3}]$\lambda 5007$/[\ion{O}{2}]$\lambda 3727$
\item[] N$_2$=[\ion{N}{2}]$\lambda 6584$/H$\alpha$
\item[] O$_3$N$_2$=([\ion{O}{3}]$\lambda 5007$/H$\beta$)/([\ion{N}{2}$\lambda 6584$/H$\alpha$]
\end{itemize}

 At $T_e$ metallicities $12+\log (\mathrm{O/H})\lesssim 8.4$, they were able to detect the temperature-sensitive [\ion{O}{3}] $\lambda 4363$ auroral line in individual galaxies. However, at higher metallicities, where the auroral line was too weak to detect in single galaxy spectra, they stacked their galaxies in bins of log[\ion{O}{2}]$\lambda$3727/H$\beta$ and log[\ion{O}{3}]$\lambda$5007/H$\beta$. Such a stacking procedure assumes no metallicity dependence on galaxy stellar mass or SFR, but only on the line ratios. In this way, they ended up with a calibration sample that extended over one decade in metallicity, covering the range $\log(\mathrm{O/H})+12=7.6-8.9$. Using this calibration sample, they (re) calibrated the relation between the $T_e$-based metallicities and  the strong emission line ratio relations given in \aref{emission line ratio relations} yielding six gas-phase metallicity diagnostics. To model the above line ratios, for each diagnostic, they fitted a polynomial of the general form $\log R=\sum_Nc_Nx^N$, where $R$ is the relevant diagnostic, $x$ is the oxygen abundance normalized to the solar value \citep[12+log(O/H)$_\odot=8.69$][]{solar}, and $c_N$ are the fitted polynomial coefficients, given in table~2 of \cite{Curti2017}.  

In this paper, we then use the method described in \cite{Curti2020} to derive a single metallicity for each of our LGRB host galaxies. We do this by finding the metallicity that minimizes the following chi-squared statistic
\begin{equation*}
    \chi^2=\displaystyle\sum_i\frac{(R_{obs,i}-R_{cal,i})^2}{\sigma_{obs}^2-\sigma_{R_{cal,i}}^2}
\end{equation*}
where, for a given diagnostic $i$, $R_{obs,i}$ is the observed line ratio, $R_{cal,i}$ is the predicted line ratio by the diagnostic for a given metallicity, and $\sigma_{obs}$ and $\sigma_{R_{cal,i}}$ are the respective observed $1\sigma$ uncertainty and intrinsic dispersion of the calibration. Similar to the method described in \aref{Metallicity_Errors}, we run a Monte Carlo Markov chain (MCMC), varying the line fluxes, sampling from a Gaussian distribution centered at the line flux measurement, and with a $\sigma$ equal to the line flux $1\sigma$ uncertainty. We then take the metallicity to be the median value from a distribution of 1000 MCMC runs, where the 1$\sigma$ uncertainty is calculated from the 16th and 85th percentiles.  This process allows \citetalias{Curti2017} to be applied on a broad range of emission line data including some line ratios which are not traditionally considered indicative of metallicity (see \citealt{KewleyReview}).

\setcounter{table}{0}
\renewcommand{\thetable}{B\arabic{table}}
\begin{table}
\footnotesize
\renewcommand{\baselinestretch}{0.80}\selectfont
\begin{center}
\begin{minipage}[H]{0.49\textwidth}
\caption{\label{xsd_data_table} Measured metallicity values}
\vspace{-0.15 cm}
\end{minipage}
\begin{tabular}{m{1.7cm}L{1.0cm}L{1.3cm}L{1.6cm}L{1.2cm}}
\hline
\hline
\multirow{2}*{Object} & Redshift & \multicolumn{2}{c}{12+log(O/H)} & Source \\
\cline{3-4}
 & (z)& \citetalias{KobulnickyKewley} $R_{23}$ & \citetalias{Curti2017} opt. & sample \\
\hline
GRB 980425  & 0.009 & 8.55 & 8.36 &  GF13\\
GRB 991208  & 0.706 & 8.05 & 8.4 &  GF13\\
GRB 000210  & 0.846 &  8.18$^{+0.52}_{-0.37}$  &  8.43$^{+0.03}_{-0.03}$  &  This work\\
GRB 010921  & 0.451 &  8.34$^{+0.26}_{-0.39}$  &  8.44$^{+0.05}_{-0.06}$  &  GF13\\
GRB 011121  & 0.362 &  8.20$^{+0.08}_{-0.06}$  &  8.59$^{+0.03}_{-0.03}$  &  GF13\\
GRB 011211  & 2.144 &  9.01$^{+0.40}_{-0.47}$  &  8.46$^{+0.25}_{-0.14}$  &  This work\\
GRB 020903  & 0.25 & 8.39 & 8.12 &  GF13\\
GRB 030329  & 0.169 & 8.12 & 8.25 &  GF13\\
GRB 031203  & 0.105 & 8.28 & 8.19 &  GF13\\
GRB 050416A  & 0.654 &  8.32$^{+0.25}_{-0.30}$  &  8.44$^{+0.03}_{-0.03}$  &  KMF+15\\
GRB 050525A  & 0.606 & 8.96 &     8.32$^{+0.42}_{-0.49}$  &  KMF+15\\
GRB 050824  & 0.828 &  8.08$^{+0.57}_{-0.23}$  &  8.09$^{+0.03}_{-0.03}$  &  KMF+15\\
GRB 050826  & 0.296 & 8.84 & 8.56 &  GF13\\
GRB 051022A  & 0.806 &  8.54$^{+0.11}_{-0.09}$  &  8.44$^{+0.02}_{-0.01}$  &  KMF+15\\
GRB 060218  & 0.033 & 8.25 & 8.22 &  GF13\\
GRB 060306  & 1.56 &  8.80$^{+0.25}_{-0.29}$  &  8.72$^{+0.26}_{-0.21}$  &  KMF+15\\
GRB 060505  & 0.089 &  8.64$^{+0.02}_{-0.22}$  &  8.59$^{+0.01}_{-0.01}$  &  GF13\\
GRB 060719  & 1.532 &  8.75$^{+0.24}_{-0.37}$  &  8.56$^{+0.20}_{-0.11}$  &  KMF+15\\
GRB 060912A  & 0.936 &  8.78$^{+0.05}_{-0.10}$  &  8.49$^{+0.04}_{-0.03}$  &  KMF+15\\
GRB 070129  & 2.338 &  8.40$^{+0.42}_{-0.56}$  &  8.39$^{+0.16}_{-0.11}$  &  KMF+15\\
GRB 070306  & 1.497 &  8.65$^{+0.10}_{-0.07}$  &  8.38$^{+0.02}_{-0.01}$  &  KMF+15\\
GRB 070612  & 0.671 & 8.17 & 8.26 &  GF13\\
GRB 070802  & 2.454 &  8.59$^{+0.19}_{-0.15}$  &  8.42$^{+0.03}_{-0.03}$  &  KMF+15\\
GRB 071021  & 2.451 &  7.89$^{+1.04}_{-0.10}$  &  8.53$^{+0.13}_{-0.07}$  &  KMF+15\\
GRB 071117  & 1.329 &  8.33$^{+0.51}_{-0.71}$  &  8.50$^{+0.20}_{-0.12}$  &  KMF+15\\
GRB 080207  & 2.086 &  8.68$^{+0.24}_{-0.20}$  &  8.47$^{+0.04}_{-0.04}$  &  KMF+15\\
GRB 080520  & 1.547 &  8.68$^{+0.10}_{-0.17}$  &  8.41$^{+0.01}_{-0.01}$  &  KMF+15\\
GRB 080605  & 1.641 &  8.57$^{+0.13}_{-0.13}$  &  8.39$^{+0.02}_{-0.02}$  &  KMF+15\\
GRB 080805  & 1.505 &  8.19$^{+0.76}_{-0.34}$  &  8.46$^{+0.06}_{-0.05}$  &  KMF+15\\
GRB 081109  & 0.979 &  8.76$^{+0.12}_{-0.15}$  &  8.62$^{+0.02}_{-0.01}$  &  KMF+15\\
GRB 081221  & 2.259 &  9.01$^{+0.06}_{-0.50}$  &  8.74$^{+0.21}_{-0.19}$  &  KMF+15\\
GRB 090407  & 1.448 &  8.73$^{+0.21}_{-0.18}$  &  8.60$^{+0.04}_{-0.03}$  &  KMF+15\\
GRB 090926B  & 1.243 &  7.93$^{+0.68}_{-0.45}$  &  8.35$^{+0.08}_{-0.07}$  &  KMF+15\\
GRB 091018  & 0.971 &  8.70$^{+0.20}_{-0.14}$  &  8.55$^{+0.08}_{-0.06}$  &  KMF+15\\
GRB 091127  & 0.49 &  8.12$^{+0.32}_{-0.16}$  &  8.37$^{+0.04}_{-0.04}$  &  KMF+15\\
GRB 100316D  & 0.059 &  8.37$^{+0.20}_{-0.06}$  &  8.31$^{+0.01}_{-0.02}$  &  KMF+15\\
GRB 100418A  & 0.623 &  8.58$^{+0.12}_{-0.15}$  &  8.45$^{+0.02}_{-0.03}$  &  KMF+15\\
GRB 100615A  & 1.398 &  7.96$^{+0.81}_{-0.33}$  &  8.33$^{+0.08}_{-0.06}$  &  KMF+15\\
GRB 100621A  & 0.543 &  8.71$^{+0.03}_{-0.07}$  &  8.35$^{+0.02}_{-0.01}$  &  KMF+15\\
GRB 100724A  & 1.289 &  8.84$^{+0.19}_{-0.76}$  &  8.56$^{+0.26}_{-0.17}$  &  KMF+15\\
GRB 100814A  & 1.439 &  8.63$^{+0.16}_{-0.23}$  &  8.26$^{+0.05}_{-0.05}$  &  KMF+15\\
GRB 100816A  & 0.805 &  7.63$^{+1.23}_{-0.41}$  &  8.48$^{+0.11}_{-0.08}$  &  KMF+15\\
GRB 110918A  & 0.984 &  8.96$^{+0.08}_{-0.13}$  &  8.73$^{+0.11}_{-0.07}$  &  KMF+15\\
GRB 120422A  & 0.283 &  8.36$^{+0.12}_{-0.08}$  &  8.49$^{+0.01}_{-0.01}$  &  KMF+15\\
GRB 120624B  & 2.197 &  8.63$^{+0.27}_{-0.20}$  &  8.43$^{+0.10}_{-0.07}$  &  KMF+15\\
GRB 120714B  & 0.398 &  8.50$^{+0.07}_{-0.15}$  &  8.42$^{+0.03}_{-0.02}$  &  KMF+15\\
GRB 120722A  & 0.959 &  8.64$^{+0.14}_{-0.13}$  &  8.42$^{+0.02}_{-0.02}$  &  KMF+15\\
GRB 120815A  & 2.359 &  8.53$^{+0.29}_{-0.29}$  &  8.27$^{+0.07}_{-0.06}$  &  KMF+15\\
GRB 121024A  & 2.301 &  8.59$^{+0.08}_{-0.09}$  &  8.29$^{+0.02}_{-0.02}$  &  KMF+15\\
GRB 130427A  & 0.34 &  8.72$^{+0.13}_{-0.34}$  &  8.56$^{+0.07}_{-0.05}$  &  KMF+15\\
GRB 130925A  & 0.348 &  8.78$^{+0.07}_{-0.06}$  &  8.56$^{+0.01}_{-0.01}$  &  KMF+15\\
GRB 131103A  & 0.596 &  8.69$^{+0.08}_{-0.46}$  &  8.36$^{+0.02}_{-0.02}$  &  KMF+15\\
GRB 131105A  & 1.685 &  8.05$^{+0.65}_{-0.55}$  &  8.52$^{+0.07}_{-0.05}$  &  KMF+15\\
GRB 131231A  & 0.643 &  8.56$^{+0.13}_{-0.09}$  &  8.32$^{+0.03}_{-0.02}$  &  KMF+15\\
GRB 140301A  & 1.416 &  8.90$^{+0.08}_{-0.10}$  &  8.68$^{+0.02}_{-0.02}$  &  KMF+15\\

\hline
\end{tabular}
\end{center}
\vspace{-0.3 cm}
\begin{minipage}[H]{0.48\textwidth} 
Objects with measured metallicity values and their associated redshifts. The third column gives the \citetalias{KobulnickyKewley} R$_{23}$ metallicities described in \aref{Host_Metallicities}, the fourth column gives the metallicity that optimizes the agreement between observed and predicted line ratios according to the six recalibrated diagnostics from \cite{Curti2017}. Further details on the \cite{Curti2020} method are given in \aref{mzd} and \aref{AppendixB}. The source sample column indicates whether the object emission line fluxes are from this work (see \aref{xsd_line_fluxes}), or have been taken from the \cite{stats_paper} (``GF13") or \cite{xshooter_survey} (``KMF+15") samples.  Duplicate objects have been removed to generate the combined sample as described in \aref{dupout}. Objects without errors do not have line fluxes with error measurement suitable to estimate errors.
\end{minipage}
\vspace{-2.7 cm}
\end{table}


\renewcommand{\thetable}{B\arabic{table}}
\begin{table}[h]
\footnotesize
\renewcommand{\baselinestretch}{0.80}\selectfont
\begin{center}
\begin{minipage}[H]{0.49\textwidth}
\caption{\label{xsd_CCM17_data_table} Measured \citetalias{Curti2017} optimal  metallicity values for LGRBs not listed in \aref{xsd_data_table}.}
\vspace{-0.15 cm}
\end{minipage}
\begin{tabular}{m{2.2cm}L{2.2cm}L{2.2cm}L{2.2cm}}
\hline
\hline
\multirow{2}*{Object} & Redshift & 12+log(O/H) \\ 
& (z) & \citetalias{Curti2017} opt. \\ 
\hline
GRB 021004 & 2.330 & 8.05$^{+0.04}_{-0.04}$ \\
GRB 051001 & 2.429 & 8.41$^{+0.08}_{-0.07}$  \\
GRB 051016B & 0.936 & 8.31$^{+0.03}_{-0.04}$ \\
GRB 051117B & 0.480 & 8.80$^{+0.15}_{-0.11}$ \\
GRB 060204B & 2.339 & 8.34$^{+0.06}_{-0.05}$ \\
GRB 060604 & 2.135 & 8.03$^{+0.06}_{-0.07}$ \\
GRB 060729 & 0.543 & 8.52$^{+0.14}_{-0.11}$ \\
GRB 060805A & 2.363 & 8.30$^{+0.05}_{-0.06}$ \\
GRB 060814 & 1.922 & 8.40$^{+0.09}_{-0.08}$ \\
GRB 061021 & 0.345 & 8.43$^{+0.05}_{-0.04}$ \\
GRB 061202 & 2.254 & 8.43$^{+0.07}_{-0.06}$ \\
GRB 070318 & 0.840 & 8.32$^{+0.03}_{-0.03}$ \\
GRB 070419B & 1.959 & 8.48$^{+0.07}_{-0.08}$ \\
GRB 080804 & 2.206 & 8.12$^{+0.12}_{-0.14}$ \\
GRB 090113 & 1.749 & 8.52$^{+0.11}_{-0.13}$ \\
GRB 090201 & 2.100 & 8.57$^{+0.09}_{-0.06}$ \\
GRB 100424A & 2.466 & 8.15$^{+0.05}_{-0.05}$ \\
GRB 100508A & 0.520 & 8.53$^{+0.08}_{-0.10}$ \\
GRB 100606A & 1.554 & 8.32$^{+0.53}_{-0.47}$ \\
GRB 100728A & 1.567 & 8.64$^{+0.25}_{-0.23}$ \\
GRB 110808A & 1.349 & 8.23$^{+0.07}_{-0.05}$ \\
GRB 111209A & 0.677 & 8.37$^{+0.05}_{-0.06}$ \\
GRB 120119A & 1.729 & 7.91$^{+1.10}_{-0.22}$ \\
GRB 140213A & 1.208 & 8.64$^{+0.28}_{-0.25}$ \\
\hline

\end{tabular}
\end{center}
\vspace{-0.3 cm}
\begin{minipage}[H]{0.48\textwidth} 
 \small Here we provide the \citetalias{Curti2017} optimal 
metallicities of objects without a \citetalias{KobulnickyKewley} metallicity and thus not listed in \aref{xsd_data_table}). \citetalias{KobulnickyKewley} requires an extensive set of emission lines (see \aref{sample}), here we have expanded our application of the \citetalias{Curti2017} optimal 
metallicity diagnostics to additional objects for which we do not have all the line fluxes required for \citetalias{KobulnickyKewley}. Regardless of whether these additional objects are included or not no redshift evolution is seen in the LGRB host metallicity distribution in the \citetalias{Curti2017} optimal metallicity diagnostic.
\end{minipage}
\vspace{-0.3 cm}
\end{table}

\renewcommand{\thetable}{B\arabic{table}}
\begin{table}
\footnotesize
\renewcommand{\baselinestretch}{0.80}\selectfont
\begin{center}
\begin{minipage}[H]{0.49\textwidth}
\caption{\label{hist_mean_table} Mean metallicity in each redshift bin.}
\end{minipage}
\begin{tabular}{L{1.6cm}L{1.2cm}L{1.0cm}L{1.2cm}L{0.0cm}L{1.0cm}L{1.2cm}}
\hline
\hline
 Redshift & \citetalias{KobulnickyKewley} & \multicolumn{2}{c}{\citetalias{Curti2017} opt.}\\
\cline{3-4}
range (z) & R$_{23}$ & matched & expanded \\
\hline
0.0~$<$~z~$<$~0.5 & 8.43 & 8.40 & 8.42 \\
0.5~$<$~z~$<$~1.0 & 8.49 & 8.42 & 8.43 \\
1,0~$<$~z~$<$~1.5 & 8.50 & 8.46 & 8.48 \\
1.5~$<$~z~$<$~2.0 & 8.51 & 8.51 & 8.44 \\
2.0~$<$~z~$<$~2.5 & 8.59 & 8.44 & 8.36 \\
\hline
0.0~$<$~z~$<$~2.5 & 8.49 & 8.43 & 8.41 \\
\hline
\end{tabular}
\end{center}
\begin{minipage}[H]{0.48\textwidth} 
\vspace{-0.3 cm}\small The average 12+log(O/H) metallicity in bins of $\Delta z =0.5$ is given for each metallicity diagnostic.  `Matched" samples are those only containing the objects with \citetalias{KobulnickyKewley} R$_{23}$ metallicity (given in \aref{xsd_data_table}) whereas `expanded" samples include additional objects with metallicities in their diagnostics (given in \aref{xsd_CCM17_data_table}).  The average metallicity of the entire {\it z}~$<$~2.5 samples is given in the bottom row.
\end{minipage}
\end{table}

In \aref{xsd_data_table} we list the metallicities we computed for objects in both the \citetalias{KobulnickyKewley} scale and using the \citetalias{Curti2017} method described above.  However the \citetalias{Curti2017} method allows us to compute the metallicities of some additional objects which lack the full line coverage needed for \citetalias{KobulnickyKewley}.  We give those metallicities in \aref{xsd_CCM17_data_table} with the warning that the incomplete line coverage introduces methodological differences depending on which lines are present.  In \aref{BF_metallicity_CDFs} left we show the resulting cumulative distribution functions, using the same redshift bins as in \aref{measured}.  In \aref{BF_metallicity_CDFs} right we show a scatter plot of our metallicities in both the \citetalias{KobulnickyKewley} and \citetalias{Curti2017} scales.  In \aref{hist_mean_table} we give the mean metallicity in each redshift bin using both the \citetalias{KobulnickyKewley} and \citetalias{Curti2017} scales, additionally dividing the latter between the sample with just the full line coverage (\aref{xsd_data_table} objects) and the expanded sample also including the \aref{xsd_CCM17_data_table} objects.  In summary, when we compare the \citetalias{Curti2017} cumulative distributions for each of the redshift bins using the K-S test, in all cases we find a p-value $>0.3$ that the distributions originate from the same parent population, with an average $D$ parameter and p-value of 0.3 and 0.5, respectively.

\setcounter{figure}{0}
\renewcommand{\thefigure}{C\arabic{figure}}
\begin{figure}[t!]
\begin{minipage}[t]{1\textwidth}
\begin{center}
\includegraphics[width=0.89\textwidth]{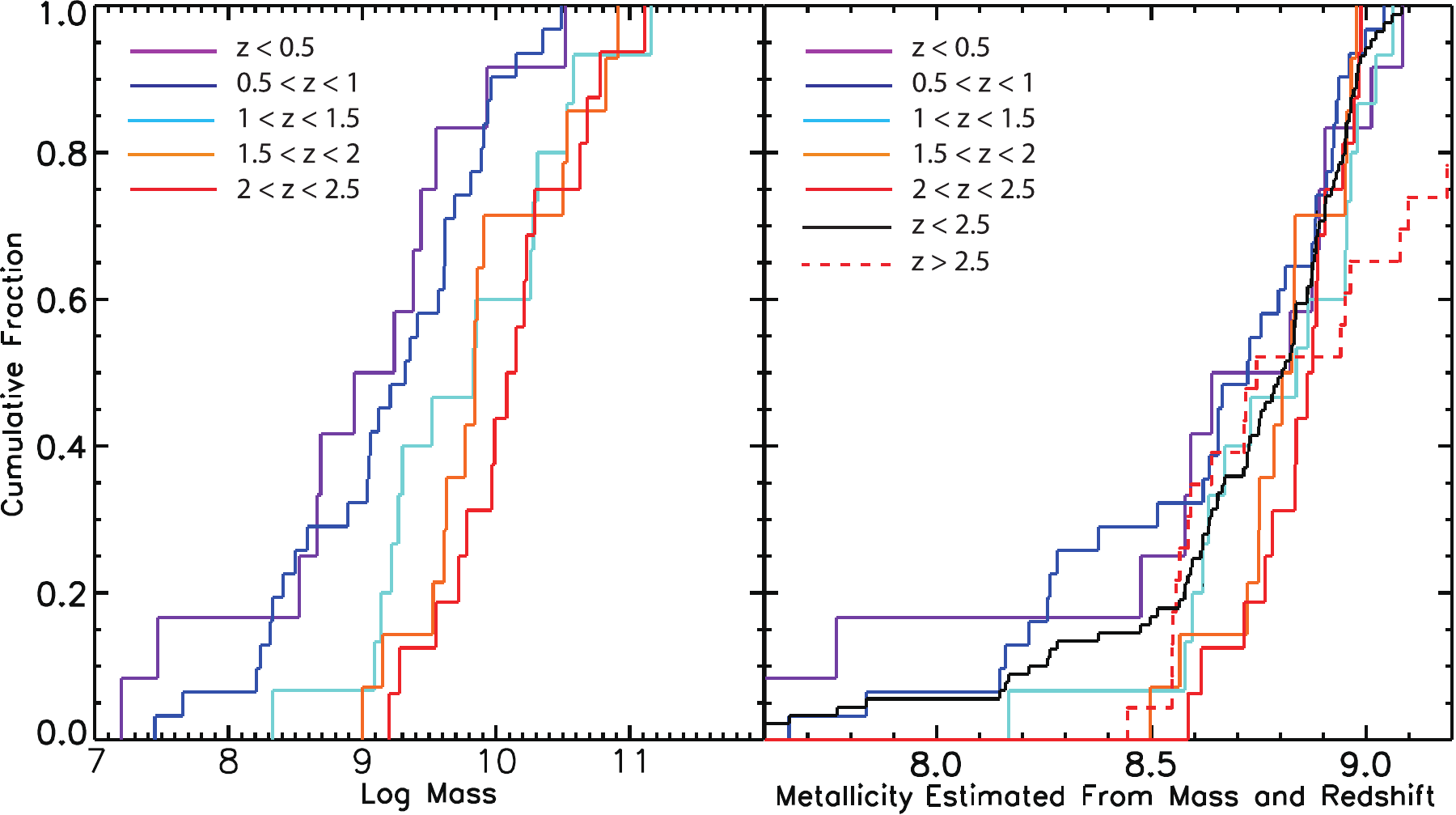}
\caption{\label{shoals_mass_mzr_metallicities}
{\it Left}: Cumulative distribution of \cite{Perley2016} SHOALS measured LGRB host masses binned by redshift. Similar to \aref{shoals_mass_step_plot_5} but no stellar mass upper limits are included in this case. {\it Right}: Cumulative distribution of estimated metallicities, based on the SHOALS stellar mass sample (plotted in left panel), which is converted to a metallicity sample using a redshift-dependent MZR from \cite{Zahid2013} (see \aref{mzd}). The data are again binned by redshift, using the same color scheme as in the left panel. This figure is comparable to \aref{shoals_metal_est_step_plot_5}, but metallicity upper limits are not included. Results from K-S statistic analysis between the distributions shown in both panels are summarized in \aref{mass_comb_step_plot_5_KS_table, metal_est_comb_step_plot_5_KS_table}.}
\end{center}
\end{minipage}
\end{figure}

\setcounter{table}{0}
\renewcommand{\thetable}{C\arabic{table}}
\begin{table}[h]
\begin{center}
\begin{minipage}[H]{0.49\textwidth}
\caption{\label{metallicity_sample_comp_KS_table}  K-S probabilities for \aref{sample_comp_comb_plot}}
\vspace{-0.15 cm}
\end{minipage}
\begin{tabular}{@{\hskip -0.05 cm}m{1.2cm}@{\hskip 0.3 cm}L{2.15cm}@{\hskip 0.3 cm}L{2.15cm}@{\hskip 0.3 cm}L{2.15cm}}
\hline
\hline
& \multicolumn{3}{c}{metallicity / mass / SFR subsamples}\\
\cline{2-4}
& Our sample & BAT6 & SHOALS \\
\hline
BAT6 &     0.87 / 0.97 / 1.00\\
SHOALS &     0.95 / 0.94 / 1.00 &     1.00 / 1.00 / 0.97\\
TOUGH &     0.92 / 1.00 / 0.17 &     1.00 / 1.00 / 0.30 &     0.98 / 0.99 / 0.10\\
\hline
\end{tabular}
\end{center}
\vspace{-0.3 cm}
\end{table}

\begin{minipage}{0.49\textwidth}
\vspace{11.9 cm}

\section{K-S tables}\label{AppendixC}

In \aref{metallicity_sample_comp_KS_table}, we give K-S test results for \aref{sample_comp_comb_plot}, providing the K-S probabilities of every line in the figure against every other line. In \aref{shoals_mass_mzr_metallicities} left and right we show a similar plot as is shown in Figures \ref{shoals_mass_step_plot_5} and \ref{shoals_metal_est_step_plot_5} respectively, but without including the stellar mass upper limits in this case, and with the inclusion of the \cite{stats_paper} objects.  The K-S test results between the various mass distributions plotted on the left and right-hand panels of \aref{shoals_mass_mzr_metallicities} are given in Tables \ref{mass_comb_step_plot_5_KS_table} and \ref{metal_est_comb_step_plot_5_KS_table}, respectively. Finally, in \aref{metal_comp_table} we provide a comparison between our measured metallicities and those estimated from the galaxy stellar mass using the typical MZR of galaxies from \cite{Zahid2013}. 

\vspace{-0 cm}
\end{minipage}

\onecolumngrid

\begin{table}[h!]
\begin{center}
\begin{minipage}[H!]{1\textwidth}
\caption{\label{mass_comb_step_plot_5_KS_table} K-S probabilities for mass distributions shown in \aref{shoals_mass_mzr_metallicities}, left panel} \vspace{-0.15 cm}
\end{minipage}
\begin{tabular}{m{2.8cm}L{1.8cm}L{1.8cm}L{1.8cm}L{1.8cm}L{1.8cm}L{1.8cm}L{1.8cm}} 
\hline
\hline
 & 0~$<$~z~$<$~0.5 (purple) & 0.5~$<$~z~$<$~1 (blue) & 1~$<$~z~$<$~1.5 (cyan) & 1.5~$<$~z~$<$~2 (orange)\\
\hline
0.5~$<$~z~$<$~1 (blue) & 0.29\\
1~$<$~z~$<$~1.5 (cyan) & 0.09 & 0.01\\
1.5~$<$~z~$<$~2 (orange) & 0.18 & 0.05 & 0.98\\
2~$<$~z~$<$~2.5 (red) & 0.02 & 0.00 & 0.42 & 0.78\\
\hline

\end{tabular}
\vspace{-2 cm}
\end{center}
\end{table}

\newpage

\begin{table*}[t]
\begin{center}
\begin{minipage}[H]{1\textwidth}
\caption{\label{metal_est_comb_step_plot_5_KS_table} K-S probabilities for MZR-based metallicity distributions shown in \aref{shoals_mass_mzr_metallicities}, right panel} \vspace{-0.15 cm}
\end{minipage}
\begin{tabular}{m{2.8cm}L{1.8cm}L{1.8cm}L{1.8cm}L{1.8cm}L{1.8cm}L{1.8cm}L{1.8cm}}
\hline
\hline
 & 0~$<$~z~$<$~0.5 (purple) & 0.5~$<$~z~$<$~1 (blue) & 1~$<$~z~$<$~1.5 (cyan) & 1.5~$<$~z~$<$~2 (orange) & 2~$<$~z~$<$~2.5 (red) & All z~$<$~2.5 (black)\\
\hline
0.5~$<$~z~$<$~1 (blue) & 0.99 (0.70) \\
1~$<$~z~$<$~1.5 (cyan) & 0.58 (0.44) & 0.26 (0.17)\\
1.5~$<$~z~$<$~2 (orange) & 0.31 (0.85) & 0.17 (0.22) & 0.66 (0.70) \\
2~$<$~z~$<$~2.5 (red) & 0.23 (0.97) & 0.06 (0.18) & 0.51 (0.42) & 0.29 (0.78) \\
All z~$<$~2.5 (black) & 0.82 (0.78) & 0.69 (0.46) & 0.58 (0.09) & 0.50 (0.81) & 0.27 (0.40) \\
z~$>$~2.5 (dashed red) & 0.36 (0.10) & 0.03 (0.00) & 0.41 (0.06) & 0.23 (0.11) & 0.19 ( 0.06) & 0.02 (0.00) \\
\hline
\end{tabular}
\end{center}
\vspace{-0.3 cm}
\begin{minipage}[H]{1\textwidth}
{First set of values given in each column are computed K-S probabilities for full sample of objects in the redshift bin of the corresponding column, whereas values in brackets give the K-S probabilities for the subsample of estimated metallicities that lie in the top half percentile of each corresponding redshift bin.}
\end{minipage}
\vspace{0.35 cm}
\newcolumntype{L}[1]{>{\centering\arraybackslash}m{#1}}
\begin{center}
\begin{minipage}[H]{1\textwidth}
\caption{\label{metal_comp_table} Measured vs.\ estimated metallicity values}
\end{minipage}
\begin{tabular}{m{2.2cm}L{1.6cm}L{1.1cm}L{0.3cm}L{1.6cm}L{1.7cm}L{1.8cm}L{1.1cm}}
\hline
\hline
\multirow{2}*{Object} & \multicolumn{2}{c}{12+log(O/H)} &   & \multicolumn{2}{c}{Metallicity difference} & Log$M_\star$ & Redshift \\
\cline{2-3}
\cline{5-6}
 & KK04 $R_{23}$ & MZR &   & Log scale & Linear scale & (Log$M_\odot)$ &  (z)\\
\hline
GRB 980425 &       8.55 &       8.59 &   &       0.04 &       0.91 &       8.53 &      0.009\\
GRB 991208 &       8.05 &       8.38 &   &       0.32 &       0.48 &       8.59 &      0.706\\
GRB 000210 &       8.18 &       8.67 &   &       0.49 &       0.32 &       9.21 &      0.846\\
GRB 010921 &       8.34 &       8.82 &   &       0.48 &       0.33 &       9.38 &      0.451\\
GRB 011121 &       8.20 &       8.90 &   &       0.70 &       0.20 &       9.55 &      0.362\\
GRB 020903 &       8.39 &       8.58 &   &       0.19 &       0.64 &       8.69 &      0.250\\
GRB 030329 &       8.12 &       7.77 &   &      -0.36 &       2.27 &       7.47 &      0.169\\
GRB 031203 &       8.28 &       8.89 &   &       0.61 &       0.24 &       9.24 &      0.105\\
GRB 050416A &       8.32 &       8.63 &   &       0.32 &       0.48 &       9.06 &      0.654\\
GRB 050525A &       8.96 &       7.84 &   &      -1.12 &      13.22 &       7.66 &      0.606\\
GRB 050824 &       8.08 &       7.66 &   &      -0.43 &       2.66 &       7.45 &      0.828\\
GRB 050826 &       8.84 &       9.01 &   &       0.17 &       0.67 &       9.93 &      0.296\\
GRB 060218 &       8.25 &       7.54 &   &      -0.71 &       5.13 &       7.20 &      0.033\\
GRB 060306 &       8.80 &       8.95 &   &       0.16 &       0.70 &      10.50 &      1.560\\
GRB 060719 &       8.75 &       8.83 &   &       0.08 &       0.84 &       9.84 &      1.532\\
GRB 060912A &       8.78 &       8.91 &   &       0.13 &       0.74 &       9.91 &      0.936\\
GRB 070129 &       8.40 &       8.88 &   &       0.47 &       0.34 &      10.15 &      2.338\\
GRB 070306 &       8.65 &       8.96 &   &       0.31 &       0.49 &      10.53 &      1.497\\
GRB 071021 &       7.89 &       8.98 &   &       1.09 &       0.08 &      10.68 &      2.451\\
GRB 080207 &       8.68 &       8.99 &   &       0.30 &       0.50 &      11.11 &      2.086\\
GRB 080605 &       8.57 &       8.95 &   &       0.38 &       0.41 &      10.53 &      1.641\\
GRB 080805 &       8.19 &       8.83 &   &       0.65 &       0.23 &       9.86 &      1.505\\
GRB 081221 &       9.01 &       8.97 &   &      -0.04 &       1.09 &      10.78 &      2.259\\
GRB 090926B &       7.93 &       8.95 &   &       1.03 &       0.09 &      10.28 &      1.243\\
GRB 091018 &       8.70 &       8.81 &   &       0.11 &       0.77 &       9.62 &      0.971\\
GRB 091127 &       8.12 &       8.48 &   &       0.36 &       0.44 &       8.66 &      0.490\\
GRB 100615A &       7.96 &       8.63 &   &       0.67 &       0.21 &       9.27 &      1.398\\
GRB 100621A &       8.71 &       8.88 &   &       0.17 &       0.67 &       9.61 &      0.543\\
GRB 100814A &       8.63 &       8.73 &   &       0.10 &       0.79 &       9.52 &      1.439\\
\hline
\end{tabular}
\end{center}
\vspace{-0.3 cm}
\begin{minipage}[H]{1\textwidth}
{Objects with both measured and estimated metallicity values.  The corresponding difference and ratio between the respective logarithmic and linear (i.e.\ nonlog) metallicities is given as well as the masses and redshifts used to calculate the estimated metallicities.  Statistics on the table above finds a mean difference between the estimated and measured metallicities of 0.23 dex with a median difference of 0.30 dex but with a standard deviation on the difference of 0.46 dex.  In raw (i.e.\ non-log) metallicities we find a median ratio of 0.496 between the measured and estimated metallicities.  Hence we conclude that while the estimated metallicities are typically higher than the measured values the offset is too random to be useful for correcting the estimated metallicities to match the measured values.  
}
\end{minipage}
\end{table*}

\clearpage

\bibliographystyle{apj_links}
\bibliography{XS_paper}
\label{end_document}

\end{document}